\documentclass[12pt]{article}

\usepackage[a4paper]{geometry} 

\usepackage[utf8]{inputenc}
\usepackage{amssymb}
\usepackage{makeidx}
\usepackage[english]{babel}
\usepackage{graphicx}
\usepackage{amsfonts,amsmath,amssymb,amsthm}
\usepackage{oldgerm}
\usepackage{relsize}

\usepackage{mathrsfs}
\usepackage[active]{srcltx}
\usepackage{verbatim}
\usepackage{enumitem} 
\usepackage[backend=biber, style=alphabetic]{biblatex}
\usepackage[toc,page]{appendix}
\usepackage{aliascnt,bbm}
\usepackage{array}

\usepackage[colorlinks = true, citecolor = blue]{hyperref}

\usepackage[textwidth=3cm,textsize=footnotesize]{todonotes}
\usepackage{xargs}
\usepackage{cellspace}

\usepackage[Symbolsmallscale]{upgreek}

\usepackage{array,multirow,makecell}
\setcellgapes{4pt}
\makegapedcells
\newcolumntype{R}[1]{>{\raggedleft\arraybackslash }b{#1}}
\newcolumntype{L}[1]{>{\raggedright\arraybackslash }b{#1}}
\newcolumntype{C}[1]{>{\centering\arraybackslash }b{#1}}

\usepackage{stmaryrd} 

\usepackage{accents}
\usepackage{dsfont}

\usepackage{aliascnt}

\usepackage{cleveref}

\makeatletter
\newtheorem{theorem}{Theorem}
\crefname{theorem}{theorem}{Theorems}
\Crefname{Theorem}{Theorem}{Theorems}

\newaliascnt{lemma}{theorem}
\newtheorem{lemma}[lemma]{Lemma}
\aliascntresetthe{lemma}
\crefname{lemma}{lemma}{lemmas}
\Crefname{Lemma}{Lemma}{Lemmas}

\newaliascnt{corollary}{theorem}

\aliascntresetthe{corollary}
\crefname{corollary}{corollary}{corollaries}
\Crefname{Corollary}{Corollary}{Corollaries}

\newaliascnt{proposition}{theorem}
\newtheorem{proposition}[proposition]{Proposition}
\aliascntresetthe{proposition}
\crefname{proposition}{proposition}{propositions}
\Crefname{Proposition}{Proposition}{Propositions}

\newaliascnt{definition}{theorem}

\aliascntresetthe{definition}
\crefname{definition}{definition}{definitions}
\Crefname{Definition}{Definition}{Definitions}

\newaliascnt{definition-proposition}{theorem}

\aliascntresetthe{definition-proposition}
\crefname{definition-proposition}{definition-proposition}{definitions-propositions}
\Crefname{Definition-Proposition}{Definition-Proposition}{Definitions-Propositions}

\newaliascnt{remark}{theorem}

\aliascntresetthe{remark}
\crefname{remark}{remark}{remarks}
\Crefname{Remark}{Remark}{Remarks}

\crefname{example}{example}{examples}
\Crefname{Example}{Example}{Examples}

\crefname{figure}{figure}{figures}
\Crefname{Figure}{Figure}{Figures}

\newtheorem{assumption}{\textbf{H}\hspace{-3pt}}
\Crefname{assumption}{\textbf{H}\hspace{-3pt}}{\textbf{H}\hspace{-3pt}}
\crefname{assumption}{\textbf{H}}{\textbf{H}}

\Crefname{probleme}{\textbf{Problem}\hspace{-3pt}}{\textbf{Problem}\hspace{-3pt}}
\crefname{probleme}{\textbf{Problem}}{\textbf{Problem}}

\Crefname{assumptionG}{\textbf{G}\hspace{-3pt}}{\textbf{G}\hspace{-3pt}}
\crefname{assumptionG}{\textbf{G}}{\textbf{G}}

\makeatother

\newcommand{\volball}{\kappa}
\newcommand{\ballUn}{\mathsf{B}}
\newcommand{\cube}{\mathsf{C}}
\newcommand{\Querm}{\mathscr{W}}
\newcommand{\VolMix}{\mathcal{V}}
\newcommand{\ensConvSet}{\mathcal{K}}
\newcommand{\EconstConvInt}{\mathsf{E}}

\def\tildR{\tilde{R}_\gamma}

\newcommand{\softOmega}{\Tilde{\Omega}}

\newcommand{\rin}{r}
\newcommand{\rext}{R}

\def\U{U}
\def\ind{\indiK}
\def\indKMY{\ind_\convSet^\lambda}
\def\m{m}

\newcommand{\bfbetaOLS}{\bfbeta^{\operatorname{OLS}}}

\def\CUnfK{\Delta_1}
\def\CDeuxfK{\Delta_2}
\def\convSetTilde{\widetilde{\convSet}}

\def\projxK{\proj{x_\convSet}{\convSetTilde}}

\def\Abor{\mathsf{A}}

\newcommandx{\functionspace}[2][1=+]{\mathbb{M}_{#1}(#2)}

\newcommandx{\VarDeux}[3][3=]{\operatorname{Var}^{#3}_{#1}\left\{#2 \right\}}

\newcommand{\1}{\mathbbm{1}}

\newcommand{\borelSet}{\mathcal{B}}

\newcommand{\LeftEqNo}{\let\veqno\@@leqno}

\newcommand{\N}{\ensuremath{\mathbb{N}}}

\newcommand{\absolute}[1]{\left\vert #1 \right\vert}
\newcommand{\abs}[1]{\left\vert #1 \right\vert}
\newcommand{\absLigne}[1]{\vert #1 \vert}
\newcommand{\tvnorm}[1]{\| #1 \|_{\mathrm{TV}}}

\newcommandx{\Vnorm}[2][1=V]{\| #2 \|_{#1}}
\newcommandx{\VnormEq}[2][1=V]{\left\| #2 \right\|_{#1}}
\newcommandx{\norm}[3][1=,2=]{\ifthenelse{\equal{#1}{}}{
    \ifthenelse{\equal{#2}{}} {\left\Vert #3 \right\Vert} {\left\Vert
        #3 \right\Vert_{#2}} } {\ifthenelse{\equal{#2}{}}{\left\Vert
        #3 \right\Vert^{#1}_{#2}}{\left\Vert #3 \right\Vert^{#1}_{#2}}
  }}
\newcommandx{\normLigne}[3][1=,2=]{\ifthenelse{\equal{#1}{}}{
    \ifthenelse{\equal{#2}{}} {\Vert #3 \Vert} {\Vert
        #3 \Vert_{#2}} } {\ifthenelse{\equal{#2}{}}{\Vert
        #3 \Vert^{#1}_{#2}}{\Vert #3 \Vert^{#1}_{#2}}
  }}
  \newcommand{\crochet}[1]{\left\langle#1 \right\rangle}
  \newcommand{\parenthese}[1]{\left(#1 \right)}
  
  \newcommand{\parentheseDeux}[1]{\left[ #1 \right]}
  \newcommand{\defEns}[1]{\left\lbrace #1 \right\rbrace }

\newcommandx{\ps}[3][1=]{\ifthenelse{\equal{#1}{}}{\left\langle#2,#3
    \right\rangle}{\left\langle#2,#3 \right\rangle_{#1}}}
\newcommandx{\psLigne}[3][1=]{\ifthenelse{\equal{#1}{}}{\langle#2,#3
    \rangle}{\langle#2,#3 \rangle_{#1}}}

\newcommandx\probaMarkovTilde[2][2=]
{\ifthenelse{\equal{#2}{}}{{\widetilde{\mathbb{P}}_{#1}}}{\widetilde{\mathbb{P}}_{#1}\left[ #2\right]}}

\newcommand{\softO}{\Tilde{\ensuremath{\mathcal O}}}

\newcommand{\couplage}[2]{\Pi(#1,#2)}

\newcommand{\plusinfty}{+\infty}

\newcounter{hypoconbis}
\newcounter{saveconbis}
\newcommand\debutH{\begin{list}
{\textbf{H\arabic{hypoconbis}}}{\usecounter{hypoconbis}}\setcounter{hypoconbis}{\value{saveconbis}}}
\newcommand\finH{\end{list}\setcounter{saveconbis}{\value{hypoconbis}}}

\def\ie{i.e.}
\def\eqsp{\;}

\newcommand{\coint}[1]{\left[#1\right)}
\newcommand{\ocint}[1]{\left(#1\right]}
\newcommand{\ooint}[1]{\left(#1\right)}
\newcommand{\ccint}[1]{\left[#1\right]}

\newcommandx{\weight}[2][2=n]{\omega_{#1,#2}}

\def\rmd{\mathrm{d}}
\newcommandx\sequence[3][2=,3=]
{\ifthenelse{\equal{#3}{}}{\ensuremath{\{ #1_{#2}\}}}{\ensuremath{\{ #1_{#2}, \eqsp #2 \in #3 \}}}}
\newcommandx{\sequencen}[2][2=n\in\N]{\ensuremath{\{ #1, \eqsp #2 \}}}
\newcommandx\sequenceDouble[4][3=,4=]
{\ifthenelse{\equal{#3}{}}{\ensuremath{\{ (#1_{#3},#2_{#3}) \}}}{\ensuremath{\{  (#1_{#3},#2_{#3}), \eqsp #3 \in #4 \}}}}
\newcommandx{\sequencenDouble}[3][3=n\in\N]{\ensuremath{\{ (#1_{n},#2_{n}), \eqsp #3 \}}}
\newcommand{\wrt}{w.r.t.}

\def\iid{i.i.d.}
\def\rme{\mathrm{e}}

\def\eg{e.g.}

\def\rset{\mathbb{R}}

\def\nset{\mathbb{N}}

\newcommandx{\CPE}[3][1=]{{\mathbb E}^{#3}_{#1}\left[#2 \right]} 
\newcommandx{\CPVar}[3][1=]{\mathrm{Var}^{#3}_{#1}\left\{ #2 \right\}}
\newcommand{\CPP}[3][]
{\ifthenelse{\equal{#1}{}}{{\mathbb P}\left(\left. #2 \, \right| #3 \right)}{{\mathbb P}_{#1}\left(\left. #2 \, \right | #3 \right)}}

\newcommandx{\osc}[2][1=]{\mathrm{osc}_{#1}(#2)}

\newcommand{\chunk}[4][]%
{\ifthenelse{\equal{#1}{}}{\ensuremath{{#2}_{#3:#4}}}{\ensuremath{#2^#1}_{#3:#4}}
}

\def\Gammabf{\mathbf{\Gamma}}

\def\bfbeta{\pmb{\beta}}
\def\betabf{\bfbeta}

\def\XE{X}
\def\ZE{Z}

\def\bfPhi{\mathbf{\Phi}}

\def\xstar{x^{\star}}

\def\Fsmall{\omega}

\def\gaStep{\gamma}

\def\RKer{R}

\def\LL{L_f}

\def\bargaStep{\bar{\gaStep}}

\newcommand{\ball}[2]{\mathrm{B}(#1,#2)}
\newcommand{\ballO}[2]{\operatorname{B}_{\text{o}}(#1,#2)}

\def\fU{f}
\def\Ul{U^{\lambdaMY}}
\def\pil{\pi^{\lambdaMY}}

\def\lambdaMY{\lambda}
\def\convSet{\mathsf{K}}
\def\indiK{\iota}

\newcommandx{\proj}[2]{\ensuremath{\operatorname{proj}_{#2}\left(  #1\right)}}
\newcommandx{\projK}[1]{\ensuremath{\operatorname{proj}_{\convSet}\left(  #1\right)}}
\newcommand{\projKw}{\ensuremath{\operatorname{proj}_{\convSet}}}

\def\Uml{U^{\lambdaMY}}

\def\piml{\pi^{\lambdaMY}}

\def\vol{\operatorname{Vol}}
\def\VolInt{\mathscr{V}}

\def\DconstConvInt{\mathsf{D}}

\def\LUl{L}

\def\linearAppl{T}
\def\eventA{\Abor}
\def\eventB{\mathsf{B}}
\def\seuil{s}

\title{Sampling from a log-concave distribution with compact support with proximal Langevin Monte Carlo}

\author{Nicolas Brosse \textsuperscript{1} \and 
Alain Durmus \textsuperscript{2} \and
 \'Eric Moulines \textsuperscript{3} \and 
 Marcelo Pereyra \textsuperscript{4}}
 
\date{}

\begin{document}
\footnotetext[1]{Centre de Math\'ematiques Appliqu\'ees, UMR 7641, Ecole Polytechnique, France. \\
 nicolas.brosse@polytechnique.edu }
\footnotetext[2]{LTCI, Telecom ParisTech
46 rue Barrault, 75634 Paris
Cedex 13, France.
 alain.durmus@telecom-paristech.fr }
 \footnotetext[3]{Centre de Math\'ematiques Appliqu\'ees, UMR 7641, Ecole Polytechnique, France. eric.moulines@polytechnique.edu }
\footnotetext[4]{School of Mathematical and Computer Sciences, Heriot-Watt University, Edinburgh, EH14 4AS, U.K.
m.pereyra@hw.ac.uk}

\maketitle

\begin{abstract}
 This paper presents a detailed theoretical analysis of the Langevin Monte Carlo sampling algorithm recently introduced in
  \cite{durmusprox:2016} when applied to log-concave probability
  distributions that are restricted to a convex body $\convSet$. This method relies on
  a regularisation procedure involving the Moreau-Yosida envelope of the
  indicator function associated with $\convSet$. Explicit
  convergence bounds in total variation norm and in Wasserstein
  distance of order $1$ are established. In particular, we show that
  the complexity of this algorithm given a first order oracle is
  polynomial in the dimension of the state space. Finally, some
  numerical experiments are presented to compare our method with
  competing MCMC approaches from the literature.
\end{abstract}


\section{Introduction}
\label{sec:intro}

Many statistical inference problems involve estimating parameters subject to constraints on the parameter space. In a Bayesian setting, these constraints define a
posterior distribution $\pi$ with bounded support. Some examples include truncated data problems which arise naturally in failure and survival time studies
\cite{klein2005survival}, ordinal data models
\cite{johnson2006ordinal}, constrained lasso and ridge regressions
\cite{celeux2012}, Latent Dirichlet Allocation \cite{blei2003latent},
and non-negative matrix factorization \cite{paisley2014bayesian}. Drawing samples
from such constrained distributions is a challenging problem that
has been investigated in many papers; see \cite{gelfand:smith:lee:1992},
\cite{pakman2014exact}, \cite{sphericalAug:2015},
\cite{Bubeck:2015}. All these works are based on efficient Markov Chain Monte
Carlo methods to approximate the posterior distribution; however, with the exception of the recent work \cite{Bubeck:2015}, these methods are not theoretically well understood and do not provide any theoretical guarantees on the estimations delivered. 

Recently a new MCMC method has been proposed in \cite{durmusprox:2016} to sample from a non-smooth log-concave probability distribution on
$\rset^d$. This method is mainly based on a carefully designed regularised version of the
target distribution $\pi$ that enjoys a number of favourable properties that are useful for MCMC simulation. In this study, we analyse the complexity of
this algorithm when applied to log-concave distributions constrained to a convex set, with a focus on complexity as the dimension of the state space increases. More
precisely, we establish explicit bounds in total variation norm and in
Wasserstein distance of order $1$ between the iterates of the Markov
kernel defined by the algorithm and the target density $\pi$.

The paper is organised as follows. \Cref{sec:presentation-MYULA}
introduces the MCMC method of \cite{durmusprox:2016}. The main
complexity result is stated in \Cref{ssec:context-contributions} and
compared to previous works on the subject. The proof of this result is
presented in \Cref{sec:dist-pi-pil} and \Cref{sec:convergence}. The
methodology is then illustrated and compared to other approaches via
experiments in \Cref{sec:numerical-exp}. Proofs are finally reported
in \Cref{sec:proofs}.


\section{The Moreau-Yosida Unadjusted Langevin Algorithm (MYULA)}
\label{sec:MYULA-result-contrib}

\subsection{Presentation of MYULA}
\label{sec:presentation-MYULA}
Let $\pi$ be a probability measure on $\rset^d$ with density \wrt~the
Lebesgue measure given for all $x \in \rset^d$ by $\pi(x) =
\rme^{-U(x)} / \int_{\rset^d} \rme^{-\U(y)} \rmd y$, where $\U :
\rset^d \to \ocint{-\infty,\plusinfty}$ is a measurable function. In the sequel, $U$ will be referred to as the potential associated with $\pi$.
Assume for the moment that $\U$ is continuously differentiable. Then, the unadjusted
Langevin algorithm (ULA) introduced in \cite{parisi:1981} (see also
\cite{roberts:tweedie-Langevin:1996}) can be used to sample from
$\pi$. This algorithm is based on the overdamped Langevin stochastic
differential equation (SDE) associated with $U$,
\begin{equation}
\label{eq:langevin}
\rmd Y_t = -\nabla \U (Y_t) \rmd t + \sqrt{2} \rmd B_t \eqsp,
\end{equation}
where $(B_t)_{t\geq0}$ is a $d$-dimensional Brownian motion.  Under
mild assumptions on $\nabla U$, this SDE has a unique strong solution
$(Y_t)_{t \geq 0}$ and defines a strong Markovian semigroup
$(P_t)_{t\geq 0}$ on $(\rset^d,\borelSet(\rset^d))$ which is ergodic
with respect to $\pi$, where $\borelSet(\rset^d)$ is the Borel
$\sigma$-field on $\rset^d$. Since simulating exact solutions of
\eqref{eq:langevin} is in general computationally impossible or very
hard, ULA considers the Euler-Maruyama discretization
associated with \eqref{eq:langevin} to approximate samples from $\pi$. Precisely, ULA constructs the discrete-time
Markov chain $(X_k)_{k \geq 0}$, started at $X_0$, given for $k \in
\nset$ by:
\begin{equation*}
\XE_{k+1} = \XE_k - \gaStep \nabla U(\XE_k) + \sqrt{2 \gaStep} \ZE_{k+1} \eqsp, 
\end{equation*}
where $\gaStep>0$ is the stepsize and $(\ZE_{k})_{k\in\nset}$ is a
sequence of \iid~standard Gaussian $d$-dimensional vectors;
the process $(X_k)_{k \geq 0}$ is used as approximate samples from $\pi$. However, the ULA algorithm cannot be
directly applied to a distribution $\pi$ restricted to a compact
convex set.
Let $\convSet \subset \rset^d$ be a convex body, \ie~a compact convex set with non-empty interior and  
$\ind_\convSet:\rset^d\to\defEns{0,\plusinfty}$ be the (convex) indicator function of $\convSet$, defined for $x\in\rset^d$ by,
\[ \ind_\convSet(x) =  
\begin{cases}
\plusinfty &  \text{if $x \notin \convSet$,} \\
 0 & \text{if $x\in\convSet$ .}
\end{cases}
\]
Let $\fU:\rset^d\to \rset$.  In this
paper we consider any probability density $\pi$ associated to a potential $U: \rset^d \to \ocint{-\infty,\plusinfty}$ of the form
\begin{equation}
\label{eq:definition-U}
\U=\fU+\ind_\convSet \eqsp,
\end{equation}
and assume that the function $\fU$ and the convex body $\convSet$ satisfy the following assumptions.
For $x\in\rset^d$ and $r>0$, denote by $\ball{x}{r}$ the closed ball of center $x$ and radius $r$: $\ball{x}{r} = \defEns{y\in\rset^d : \norm{y-x}\leq r}$.
\begin{assumption}
\label{assumption:form-potential}
\begin{enumerate}[label=(\roman*)]
\item 
\label{item:assum-fU-convex} 
$\fU$ is convex.
\item 
\label{item:assum-fU-grad}
$\fU$ is continuously differentiable on $\rset^d$ and gradient Lipschitz with Lipschitz constant $\LL$, \ie~for all $x,y \in \rset^d$
\begin{equation}
\label{eq:gradient-Lip}
\norm{\nabla \fU(x) - \nabla \fU(y)} \leq \LL \norm{x-y} \eqsp.
\end{equation}
\end{enumerate}
\end{assumption}

\begin{assumption}
\label{assumption:K} 
There exist $\rin, \rext >0$, $\rin\leq\rext$, such that,
\begin{equation*}
\ball{0}{\rin} \subset \convSet \subset \ball{0}{\rext} \eqsp .
\end{equation*}
\end{assumption}

To apply ULA, \cite{durmusprox:2016} suggested to carefully regularize $\U$ in such a way that 1) the convexity of $\U$ is preserved (this property is key to the theoretical analysis of the algorithm), 2) the regularisation of $\U$ is continuously differentiable and gradient Lipschitz (this regularity property is key to the algorithm's stability), and 3) the resulting approximation  is close to $\pi$ (\eg~in total variation norm). The tool used to construct such an approximation is the Moreau-Yosida envelope of $\ind_\convSet$, $\indKMY:\rset^d \to \rset_{+}$ defined for $x\in\rset^d$ (see \eg~\cite[Chapter 1 Section G]{rockafellar:wets:1998}) by,
\begin{equation}
\label{eq:def-indKMY}
\indKMY(x) = \inf_{y\in\rset^d} \parenthese{\ind_\convSet(y) + (2\lambdaMY)^{-1}\norm[2]{x-y}} = (2\lambdaMY)^{-1}\norm[2]{x-\projK{x}} \eqsp ,
\end{equation}
where $\lambdaMY>0$ is a regularization parameter and $\projKw$ is the projection onto $\convSet$.
By \cite[Example 10.32, Theorem 9.18]{rockafellar:wets:1998}, the function $\indKMY$ is convex and continuously differentiable with gradient given  for all $x \in \rset^d$ by:
\begin{equation}
\label{eq:def-grad-indKMY}
\nabla \indKMY(x) = \lambdaMY^{-1}(x-\projK{x}) \eqsp.
\end{equation}
Moreover, \cite[Proposition 12.19]{rockafellar:wets:1998} implies that $\indKMY$ is $\lambdaMY^{-1}$-gradient Lipschitz: for all $x,y \in \rset^d$,
\begin{equation}
\label{eq:lip-grad-indKMY}
\norm{\nabla \indKMY(x) - \nabla \indKMY(y)} \leq \lambdaMY^{-1} \norm{x-y} \eqsp.
\end{equation}
Adding $\fU$ to $\indKMY$ under \Cref{assumption:form-potential} leads to the regularization $\Uml:\rset^d \to \rset$ of the potential $U$ defined for all $x \in \rset^d$ by
\begin{equation}\label{eq:def-Uml}
\Uml(x) = \fU(x) + \indKMY(x) \eqsp.
\end{equation} 
The following lemma shows that the probability measure $\piml$ on $\rset^d$, with density with respect to the Lebesgue measure, also denoted by $\piml$ and given for all $x \in \rset^d$ by 
\begin{equation}\label{eq:def-pilambda}
\pi^\lambda (x)= \frac{\rme^{-U^\lambda(x)}}{\int_{\rset^d} \rme^{-U^\lambda (s)} \rmd s} \eqsp,
\end{equation}
is well defined. It also shows that $\Uml$ has a minimizer $\xstar\in\rset^d$, a fact that will be used in \Cref{sec:convergence}.

\begin{lemma}\label{lemma:pilambda-defined}
Assume \Cref{assumption:form-potential}-\ref{item:assum-fU-convex} and \Cref{assumption:K}. For all $\lambdaMY >0$ ,
\begin{enumerate}[label=\alph*)]
\item 
\label{item:Ulambdaminimizer}
$\Uml$ has a minimizer $\xstar\in\rset^d$, \ie~for all $x\in\rset^d$, $\Uml(x) \geq \Uml(\xstar)$.
\item
\label{item:pilambda-defined}
$\rme^{-\Uml}$ defines a proper density of a probability measure on $\rset^d$, \ie 
\begin{equation*}
0 < \int_{\rset^d} \rme^{-\Uml(y)}\rmd y < \plusinfty \eqsp.
\end{equation*}
\end{enumerate}
\end{lemma}

\begin{proof}
Note that \cite[Proposition 1]{durmusprox:2016} provides a proof in a more general case. Given the specific form of $\Uml$, a short and self-contained proof can be found in \Cref{sec:proof-lemma-pilambda-defined}.
\end{proof}
Under \Cref{assumption:form-potential}, for all $\lambdaMY >0$, $\pi^\lambda$ is log-concave and $U^\lambda$ is continuously differentiable by \eqref{eq:def-grad-indKMY}, with $\nabla U^{\lambda}$ given for all $x \in \rset^d$ by 
\begin{equation}
\label{eq:definition-grad-prox_U}
   \nabla U^\lambda (x)=-\nabla \log \pi^\lambda(x) =  \nabla f(x) +\lambdaMY^{-1}(x-\projK{x}) \eqsp.
\end{equation}
In addition, by \eqref{eq:lip-grad-indKMY}, $\nabla \Uml$ is Lipschitz
with constant $\LUl \leq \LL + \lambda^{-1}$.  Since $\Uml$ is
continuously differentiable, ULA is well defined. The algorithm proposed in \cite{durmusprox:2016}
then proceeds by using the
Euler-Maruyama discretization of the Langevin equation associated with
$\Ul$, with $\pi^\lambda$ as proxy, to generate approximate samples from $\pi$. Precisely, it uses the Markov chain $(\XE_k)_{k\in\nset}$, started at $X_0$, given for all $k \in \nset$ by
\begin{equation}
  \label{eq:def-MYRULA}
  \XE_{k+1} = (1- \tfrac{\gaStep}{\lambda})\XE_{k} - \gaStep \nabla \fU(\XE_k)  + \tfrac{\gaStep}{\lambda}\projK{\XE_k} +\sqrt{2 \gaStep} \ZE_{k+1} \eqsp,
\end{equation}
where $(\ZE_{k})_{k\in\nset}$ is a sequence of \iid~standard Gaussian $d$-dimensional vectors and $\gaStep>0$ is the stepsize. 
Note that one iteration \eqref{eq:def-MYRULA} requires a projection onto the convex body $\convSet$ and the evaluation of $\nabla f$.
The kernel of the homogeneous Markov chain defined by \eqref{eq:def-MYRULA} is given for $x\in\rset^d$ and $\Abor \in \borelSet(\rset^d)$ by,
\begin{equation}
\label{eq:definition_R_kernel}
\RKer_{\gaStep}(x ,\Abor) = (4 \uppi \gaStep)^{-d/2}\int_{\Abor}\exp \parenthese{- (4 \gaStep)^{-1}\norm[2]{y-x+\gaStep \nabla \Ul(x)}} \rmd y \eqsp,
\end{equation}
where $\Uml$ is defined in \eqref{eq:def-Uml}. 
Since the target density for the Markov chain \eqref{eq:def-MYRULA} is the regularized measure $\pil$ and not $\pi$, the algorithm is named the Moreau-Yosida regularized Unadjusted Langevin Algorithm (MYULA).

\subsection{Context and contributions}
\label{ssec:context-contributions}

The total variation distance between two probability measures $\mu$ and $\nu$ is defined by $\tvnorm{\mu-\nu} = 2 \sup_{\Abor \in \borelSet(\rset^d)} \absolute{\mu(\Abor)-\nu(\Abor)}$.
Let $\phi,\psi : \rset_+ \to \rset_+$. Denote by $\phi = \softO(\psi)$ or $\phi = \softOmega(\psi)$ if there exist $C,c \geq 0$ such that for all $t \in \rset_+$ $\phi(t) \leq C\psi(t) (\log t)^c$ or $\phi(t) \geq C\psi(t) (\log t)^c$ respectively.
Our main result is the following:

\begin{theorem}\label{thm:tv-O-dimension}
Assume  \Cref{assumption:form-potential} and \Cref{assumption:K}. 
For all $\varepsilon >0$ and $x\in\rset^d$, there exist $\lambdaMY >0$ and $\gaStep \in \ooint{0,\lambdaMY (1+\LL^2 \lambda^2)^{-1}}$ such that,
\begin{equation*}
\tvnorm{\delta_x  \RKer_{\gaStep}^n-\pi} \leq \varepsilon \quad \text{for} \quad n = \softOmega(d^5) \eqsp,
\end{equation*}
where $\RKer_{\gaStep}$ is defined in \eqref{eq:definition_R_kernel}.
\end{theorem}
The proof of \Cref{thm:tv-O-dimension} follows from combining \Cref{theo:convergence_TV_dec-stepsize-StV} and \Cref{propo:finite-measure-MY} below. 
Note that 
these two results imply explicit bounds between $\RKer_ {\gaStep}^n$ and $\pi$ for all $n \in \nset$ and $\gaStep >0$.

The problem of sampling from a probability measure restricted to a convex compact
support has been investigated in several works, mainly in the fields of
theoretical computer science and Bayesian statistics. In computer
science, a line of works starting with \cite{dyer1991computing} has
studied the convergence of the ball walk and the hit-and-run algorithm
towards the uniform density on a convex body $\convSet$, or more
generally to a log-concave density. The best complexity result is
achieved by \cite[Theorem 2.1]{Lovasz:2007} who establishes a mixing
time for these two algorithms of order $\softO(d^4)$.
However, observe that contrary to \Cref{thm:tv-O-dimension}, this result assumes that $\pi$
is in near-isotropic position, \ie~there exists $C\in \rset_+^*$ such
that for all $u \in \rset^d$, $\norm{u} =1$,
\begin{equation}\label{eq:condition-isotropic}
 C^{-1} \leq \int_{\rset^d} \ps{u}{x}^2 \pi(\rmd x) \leq C \eqsp.
\end{equation}
Note that \cite[Section 2.5]{Lovasz:2007} gives also an algorithm of complexity
$\softO(d^5)$ which provides an invertible linear map $\linearAppl$ of
$\rset^d$ such that the measure $\pi_{\linearAppl}$ defined for all $\eventA \in \borelSet(\rset^d)$ by
\begin{equation*}
  \pi_{\linearAppl}(\eventA) =   \pi( \linearAppl^{-1}(\eventA)) \eqsp,
\end{equation*}
is log-concave and near-isotropic. Also note that, unlike our
method, each iteration of the ball walk or the hit-and-run
algorithm requires a call to a zero-order oracle, which given $x \in
\rset^d$, returns the value
$U(x)$. 
MYULA does not require to fulfill the condition \eqref{eq:condition-isotropic} and is thus dispensed of preprocessing step. However, MYULA needs a first-order oracle which returns the value $\nabla f(x)$ for $x\in\rset^d$. 


As emphasized in the introdution, probability distributions with convex
compact supports or more generally with constrained parameters arise
naturally in Bayesian statistics. \cite{gelfand:smith:lee:1992}
includes many examples of such problems and suggests to use a Gibbs
sampler, see
also \cite{rodriguez2004efficient}. \cite[Chapter 6]{chen2012monte} addresses the subject with
the additional difficulty of computing normalizing
constants. Recently, \cite{pakman2014exact} adapted the Hamiltonian
Monte Carlo method to sample from a truncated multivariate gaussian,
and \cite{sphericalAug:2015} suggested a new approach which consists in mapping
the constrained domain to a sphere in an augmented space. However, these methods are not well understood from a theoretical viewpoint, and
do not provide any theoretical guarantees for the estimations delivered.


Concerning the ULA algorithm, when $U$ is continuously differentiable,
the first explicit convergence bounds have been obtained by
\cite{dalalyan2016theoretical}, \cite{durmus2015non}, \cite{durmus:highdimULA2016}.
In the
constrained case $U=f+\indiK_\convSet$, \cite{Bubeck:2015} suggests a
projection step in ULA \ie~to consider the Markov chain
$(\tilde{\XE}_k)_{k \geq 0}$, defined for all $k \in \nset$ by
\begin{equation}\label{eq:def-projULA}
\tilde{\XE}_{k+1} = \projK{\tilde{\XE}_k - \gaStep \nabla U(\tilde{\XE}_k) + \sqrt{2 \gaStep} \ZE_{k+1}}  \eqsp.
\end{equation}
with $\tilde{\XE}_0=0$. 
This method is referred to as the Projected Langevin Monte Carlo (PLMC)
algorithm. As in MYULA, one iteration of PLMC requires a
projection onto $\convSet$ and an evaluation of $\nabla f$.
Let $\tildR$ be the Markov kernel defined by \eqref{eq:def-projULA}.
\cite{Bubeck:2015} proved that for all $\varepsilon>0$, $\tvnorm{\delta_0 \tildR^n - \pi} \leq \varepsilon$ for $n=\softOmega(d^7)$ if $\pi$ is the uniform density on $\convSet$ and $n=\softOmega(d^{12})$ if $\pi$ is a log-concave density.  
\Cref{thm:tv-O-dimension} improves these bounds for the MYULA algorithm. Note however that the iterations of PLMC stay within the constraint set $\convSet$ and this property can be useful in some specific problems. Nevertheless, there is a wide range of settings where this property is not particularly beneficial, for example in the case of the computation of volumes discussed in \Cref{sec:numerical-exp}, or in Bayesian model selection where it is necessary to estimate marginal likelihoods.

\section{Distance between $\pi$ and $\pil$}
\label{sec:dist-pi-pil}

In this section, we derive bounds between $\pi$ and $\pil$ in total
variation and in Wasserstein distance (recall that $\pi$ is associated
with a potential of the form \eqref{eq:definition-U} and $\piml$ is
given by \eqref{eq:def-pilambda}). It is shown that the approximation
error in both distances can be made arbitrarily small by adjusting the
regularisation parameter $\lambdaMY$.

The main quantity of interest to analyze the distance between $\pi$ and $\pil$ will appear to be the integral of $x \mapsto \rme^{-(2  \lambdaMY)^{-1} \norm[2]{x- \proj{x}{\convSet}}}$ over $\rset^d$. This constant is linked to useful notions borrowed from the field of convex geometry \cite[Proposition 3]{kampf:2009}. Indeed, Fubini's theorem gives the following equality:
\begin{align}
\nonumber
\int_{\rset^d} \rme^{-(2  \lambdaMY)^{-1} \norm[2]{x- \proj{x}{\convSet}}} \rmd x &= \int_{\rset_+} \int_{\rset^d} \1_{ \coint{\norm{x-\projK x},\plusinfty}}(t) \lambda^{-1} t\rme^{-t^2/(2 \lambdaMY)} \rmd x \rmd t \eqsp, \\
\label{eq:kampf}
&= \int_{\rset_+} \vol\parenthese{\convSet + \ball{0}{t}} \lambda^{-1} t\rme^{-t^2/(2 \lambdaMY)} \rmd t \eqsp,
\end{align}
where $\eventA+\eventB$ is the Minkowski sum of $\eventA,\eventB \subset \rset^d$, \ie~$\eventA+\eventB=\defEns{x+y : x\in \eventA, y\in \eventB}$, and we have used in the last line that for all $t \in \rset_+$, $\convSet + \ball{0}{t} = \{ x \in \rset^d \, : \, \norm{x-\projK x} \leq t\}$.
It turns out that $t\mapsto \vol\parenthese{\convSet + \ball{0}{t}}$ on $\rset_+$ is a polynomial.
More precisely, Steiner's formula states that for all $t \geq 0$,
\begin{equation}
\label{eq:Steiner_formula}
\vol(\convSet + \ball{0}{t}) = \sum_{i=0}^{d} t^{i} \volball_{i} \VolInt_{d-i}(\convSet) \eqsp,
\end{equation}
where $\{\VolInt_i(\convSet)\}_{0 \leq i \leq d}$ are the intrinsic volumes of $\convSet$, $\volball_i$ denotes the volume of the unit ball in $\rset^i$, \ie~
\begin{equation}
  \label{eq:vol_ball}
\volball_i = \uppi^{i/2} / \Gammabf(1+i/2) \eqsp,
\end{equation}
 and $\Gammabf: \rset_+^* \to \rset_+^*$ is the Gamma function.
We refer to \cite[Chapter 4.2]{schneider:2013} for this result and an introduction to this topic.
Combining \eqref{eq:kampf} and \eqref{eq:Steiner_formula} gives:
 \begin{equation}
 \label{eq:Wills}
   \int_{\rset^d} \rme^{-(2  \lambdaMY)^{-1} \norm[2]{x- \proj{x}{\convSet}}} \rmd x = \sum_{i=0}^{d} \VolInt_i(\convSet) (2\uppi\lambda)^{(d-i)/2} \eqsp.
 \end{equation}
This expression will provide a precise analysis of the distance in total variation and Wasserstein distance between $\pi$ and $\piml$, in particular when $\pi$ is the uniform density on $\convSet$. However, in more general cases, an additional assumption on the relation between $f$ and $\convSet$ is necessary to bound the distance between $\pi$ and $\piml$. Under \Cref{assumption:form-potential}-\ref{item:assum-fU-convex} and \Cref{assumption:K}, $f$ has a minimum $x_\convSet$ on $\convSet$. Define
\begin{equation}\label{eq:def-convSetTilde}
\convSetTilde = \defEns{x \in \convSet \eqsp | \eqsp \ball{x}{\rin} \subset \convSet} \eqsp .
\end{equation}
$\convSetTilde$ has the following property.

\begin{lemma}\label{lemma:KTilde}
Assume \Cref{assumption:K}.
$\convSetTilde$ is a non-empty convex compact set.
\end{lemma}

\begin{proof}
The proof is postponed to \Cref{sec:proof-lemmaKTilde}.
\end{proof}


\begin{assumption}
\label{assumption:fK} 
\begin{enumerate}[label=(\roman*)]
\item 
\label{item:assum-fK-1} 
There exists $\CUnfK >0$ such that $\exp\parenthese{\inf_{\convSet^{c}} (f) - \max_\convSet (f)} \geq \CUnfK$.
\item 
\label{item:assum-fK-2}
There exists $\CDeuxfK \geq 0$ such that $0 \leq f(\proj{x_\convSet}{\convSetTilde}) -f(x_\convSet) \leq \CDeuxfK$.
\end{enumerate}
\end{assumption}

Under \Cref{assumption:fK}-\ref{item:assum-fK-1}, the application of Steiner's formula is possible and reveals the precise dependence of the bounds with respect to the intrinsic volumes of $\convSet$. 
A complementary view is possible under \Cref{assumption:fK}-\ref{item:assum-fK-2}. The obtained bounds are less precise regarding $\convSet$ but more robust with respect to $f$.
Note that if $x_\convSet \in \convSetTilde$, $\CDeuxfK$ can be chosen equal to $0$.
On the other hand, if $f$ is assumed to be $\ell$-Lipschitz inside $\convSet$, $\CDeuxfK$ is less than $\ell \rext$.

\begin{proposition}\label{propo:finite-measure-MY}
Assume \Cref{assumption:form-potential}-\ref{item:assum-fU-convex} and \Cref{assumption:K}.
\begin{enumerate}[label=\alph*)]
\item 
\label{item:propo:dist_TV_MY_1}
Assume \Cref{assumption:fK}-\ref{item:assum-fK-1}.
For all $\lambdaMY >0$, 
\begin{equation}
\label{eq:propo:dist_TV_MY_1}
\tvnorm{\piml-\pi} \leq 2 \parenthese{1+ \CUnfK \DconstConvInt(\convSet,\lambdaMY)^{-1}}^{-1} \eqsp,
\end{equation}
where,
\begin{equation}\label{eq:def-DKlambda}
\DconstConvInt(\convSet,\lambdaMY) =   (\vol{\convSet})^{-1}\sum_{i=0}^{d-1} (2\uppi\lambdaMY)^{(d-i)/2} \VolInt_i(\convSet) \eqsp,
\end{equation}
and $\VolInt_i(\convSet)$ are defined in \eqref{eq:Steiner_formula}.
\item
\label{item:propo:dist_TV_MY_2}
In addition, assuming \Cref{assumption:fK}-\ref{item:assum-fK-1}, for all $\lambdaMY \in\ooint{0,(2\uppi)^{-1}(\rin/d)^2}$, 
\begin{equation}\label{eq:propo:dist_TV_MY_2}
\tvnorm{\piml-\pi} \leq  2^{3/2} \CUnfK^{-1} (\uppi\lambda)^{1/2}d\rin^{-1} \eqsp .
\end{equation}
\item
\label{item:propo:dist_TV_MY_3}
Assume \Cref{assumption:fK}-\ref{item:assum-fK-2}.
For all $\lambda\in\ocint{0, 16^{-1}(\rin/d)^2}$,
\begin{equation}\label{eq:propo:dist_TV_MY_3}
\tvnorm{\piml-\pi} \leq 
(4/\rin)\exp\parenthese{4\lambda\parenthese{\CDeuxfK/\rin}^2}
\defEns{\sqrt{\lambda}(d+\CDeuxfK) + (2\lambda\CDeuxfK)/\rin} \eqsp .
\end{equation}
\end{enumerate}
\end{proposition}

\begin{proof}
The proof is postponed to \Cref{sec:proof-pi-pilambda-tv}.
\end{proof}

In the particular case where $\fU=0$ and $\pi$ is the uniform density
on $\convSet$, $\CUnfK$ equals $1$ and the inequality \eqref{eq:propo:dist_TV_MY_1} is in fact
an equality.  The dependence of the upper bound in
\eqref{eq:propo:dist_TV_MY_1} \wrt~to $\lambdaMY, d, \rin$ is
sharp. Indeed, for the cube $\cube$ of side $c$,
$\DconstConvInt(\cube,\lambdaMY)$ can be explicitly
computed. \cite[Theorem 4.2.1]{klain1997introduction} gives for
$i\in\defEns{0,\ldots,d}$, $\VolInt_i(\cube)=\binom{d}{i}c^i$, which
implies:
\begin{align*}
\DconstConvInt(\cube,\lambdaMY) &= \parenthese{1+c^{-1}\sqrt{2\uppi\lambda}}^d -1 \eqsp , \\
\tvnorm{\piml-\pi} &= 2 \defEns{1-\parenthese{1+c^{-1}\sqrt{2\uppi\lambda}}^{-d}} \eqsp, \text{ for } U = \indiK_{\cube} \eqsp.
\end{align*}

For two probability measures $\mu$ and $\nu$ on $\borelSet(\rset^d)$,
the Wasserstein distance of order $p\in\nset^{*}$ between $\mu$ and
$\nu$ is defined by
\begin{equation*}
W_p(\mu,\nu) = \left( \inf_{\zeta \in \couplage{\mu}{\nu}} \int_{\rset^d \times \rset^d} \norm[p]{x-y}\rmd \zeta (x,y)\right)^{1/p} \eqsp,
\end{equation*}
where $\couplage{\mu}{\nu}$ is the set of transference plans of $\mu$ and $\nu$. $\zeta$ is a transference plan of $\mu$ and $\nu$ if it is a probability measure on $(\rset^d \times \rset^d, \mathcal{B}(\rset^d \times \rset^d) )$  such that for all  $\Abor \in \borelSet(\rset^d)$, $\zeta(\Abor \times \rset^d) = \mu(\Abor)$ and $\zeta(\rset^d \times \Abor) = \nu(\Abor)$.

\begin{proposition}\label{propo:wasserstein-bounds}
Assume \Cref{assumption:form-potential}-\ref{item:assum-fU-convex} and \Cref{assumption:K}.
\begin{enumerate}[label=\alph*)]
\item 
\label{item:propo:wasserstein_1}
Assume \Cref{assumption:fK}-\ref{item:assum-fK-1}.
For all $\lambdaMY>0$,
\begin{equation*}
W_1(\pi,\pil) \leq \CUnfK^{-1} \EconstConvInt(\convSet,\lambdaMY, \rext) \eqsp ,
\end{equation*}
where
\begin{equation*}
\EconstConvInt(\convSet,\lambdaMY, \rext) = (\vol(\convSet))^{-1}\sum_{i=0}^{d-1} \VolInt_i(\convSet) \parenthese{2\uppi\lambda}^{(d-i)/2} 
\defEns{2\rext + \left[\lambda(d-i+2)\right]^{1/2} } \eqsp ,
\end{equation*}
and $\VolInt_i(\convSet)$ are defined in \eqref{eq:Steiner_formula}.
\item 
\label{item:propo:wasserstein_2}
In addition, assuming \Cref{assumption:fK}-\ref{item:assum-fK-1}, for all $\lambdaMY \in\ooint{0,(2\pi)^{-1}d^{-2}\rin^2}$,
\begin{equation*}
W_1(\pi,\pil) \leq \CUnfK^{-1} (2\uppi\lambda)^{1/2}d\rin^{-1} 
\parenthese{2\rext+\rin\parenthese{3/(2d\uppi)}^{1/2}} \eqsp .
\end{equation*}
\item 
\label{item:propo:wasserstein_3}
Assume \Cref{assumption:fK}-\ref{item:assum-fK-2}.
For all $\lambda\in\ocint{0, 16^{-1}(\rin/d)^2}$,
\begin{equation*}
W_1(\pi,\pil) \leq
4\exp\parenthese{4\lambda\parenthese{\CDeuxfK/\rin}^2}
\defEns{\sqrt{\lambda}(d+\CDeuxfK)(\rext/\rin) + (2\lambda\CDeuxfK\rext)/\rin^2 + \sqrt{\pi \lambda}} \eqsp .
\end{equation*}
\end{enumerate}
\end{proposition}

\begin{proof}
The proof is postponed to \Cref{sec:proof-pi-pilambda-wasserstein}.
\end{proof}

Note that the bounds in Wasserstein distance between $\pi$ and $\pil$ are roughly similar to those obtained in total variation norm.

\section{Convergence analysis of MYULA}
\label{sec:convergence}
We now analyse the convergence of the Markov kernel $\RKer_{\gaStep}$, given by \eqref{eq:definition_R_kernel}, to the target density $\pil$ defined in \eqref{eq:def-pilambda}. 
For $x\in\rset^d$ and $n\in\nset$, explicit bounds in total variation norm and in Wasserstein distance between $\delta_x  \RKer_{\gaStep}^n$ and $\pil$ are provided in \Cref{theo:convergence_TV_dec-stepsize-StV} and \Cref{prop:wasserstein-epsilon}. 
Because of the regularisation procedure performed in \Cref{sec:presentation-MYULA}, the convergence analysis of MYULA \eqref{eq:def-MYRULA} is an application of results of \cite{durmus2015non} and \cite{durmus:highdimULA2016}. 

\subsection{Convergence in total variation norm}
\label{subsec:convergence-tv}

Define $\Fsmall:\rset_+ \to \rset_+$  for all $r\geq 0$ by
\begin{equation}
\label{eq:Fsmall}
\Fsmall(r) = r^{2}/ \defEns{2\bfPhi^{-1}(3/4)}^{2} \eqsp,
\end{equation}
where $\bfPhi(x)=(2\uppi)^{-1/2} \int_{-\infty}^x \rme^{-t^2/2} \rmd t$.

\begin{proposition}
\label{theo:convergence_TV_dec-stepsize-StV}
Assume  \Cref{assumption:form-potential} and \Cref{assumption:K}.
Let $\lambdaMY>0$, $\LUl$ be the Lipschitz constant of $\nabla \Uml$ defined in \eqref{eq:def-Uml} and $\bargaStep \in \ooint{0,\lambdaMY^{-1} \LUl^{-2}}$. 
Then for all $\varepsilon >0$ and $x\in\rset^d$, we get:
\begin{equation}\label{eq:tvnorm-RKerPilambda}
\tvnorm{\delta_x  \RKer_{\gaStep}^n-\pil} \leq \varepsilon \eqsp,
\end{equation}
provided that  $n > T \gaStep^{-1}$ with
\begin{subequations}
\begin{align}
T &= \parenthese{ \log\{ A_2(x) \}-\log(\varepsilon/2)} \Big/(- \log(\kappa)) \eqsp , \\
 \gaStep &\leq \frac{-d+\sqrt{d^2 +(2/3) A_1(x) \varepsilon^2 (\LUl^2T)^{-1} }}{2 A_1(x)/3} \wedge \bargaStep  \eqsp ,
\end{align}
\label{eq:precision_convex}
\end{subequations}
where
\begin{align*}
  A_1(x) & = L^2 \parenthese{\norm[2]{x-\xstar} + 2(d + 8\lambdaMY^{-1} \rext^2)\rme^{\gaStep(\lambdaMY^{-1}-\bargaStep \LUl^2)}(\lambdaMY^{-1}-\bargaStep \LUl^2)^{-1}} \eqsp ,\\
 \log(\kappa) 
 &= - \log(2)(4\lambdaMY)^{-1} \parentheseDeux{\log \defEns{ \parenthese{1+ \rme^{(8\lambdaMY)^{-1} \Fsmall\defEns{\max(1,4\rext)}}}\parenthese{1+\max(1,4\rext)}} +\log(2) }^{-1}   \eqsp, \\
A_2(x) &= 6 + 2^{3/2}\parenthese{d\lambdaMY + 8\rext^2}^{1/2} + 2(A_1(x)/\LUl^2)^{1/2}
 \eqsp,
\end{align*}
and $x^{\star}$ is a minimizer of $\Ul$.
\end{proposition}

\begin{proof}
To apply \cite[Theorem 21]{durmus2015non}, it is sufficient to check the assumption \cite[H3]{durmus2015non}, \ie~there exist $\tilde{R} \geq 0$ and $m >0$ such that for all $x,y \in \rset^d$, $\norm{x-y} \geq \tilde{R}$,
\begin{equation}
\label{eq:1}
  \ps{\nabla \Ul(x)-\nabla \Ul(y)}{x-y} \geq
m \norm[2]{x-y} \eqsp.
\end{equation}
By \eqref{eq:def-grad-indKMY} and the Cauchy-Schwarz inequality, we have:
\begin{equation*}
\ps{\nabla \indKMY(x)-\nabla \indKMY(y)}{x-y} \geq
\lambdaMY^{-1}\parenthese{\norm[2]{x-y}-2\defEns{ \sup_{z \in \convSet} \norm{z}} \norm{x-y}}  \eqsp,
\end{equation*}
which implies under \Cref{assumption:form-potential}-\ref{item:assum-fU-convex} and \Cref{assumption:K} that \eqref{eq:1} holds for $\tilde{R} = 4R$ and $m =(2\lambdaMY)^{-1}$.
\end{proof}

Combining \Cref{propo:finite-measure-MY} and \Cref{theo:convergence_TV_dec-stepsize-StV} determines the stepsize $\gamma$ and the number of samples $n$ to get $\tvnorm{\delta_{\xstar}  \RKer_{\gaStep}^n-\pi} \leq \varepsilon$. 
$\lambdaMY$ is chosen of order $\varepsilon^2 \rin^2 d^{-2}\CUnfK^{2}$ under \Cref{assumption:fK}-\ref{item:assum-fK-1} and $\varepsilon^2 \rin^2 \min(d^{-2}, \CDeuxfK^{-2})$ under \Cref{assumption:fK}-\ref{item:assum-fK-2}.
The orders of magnitude of $n$ in $d,\varepsilon,\rext,\rin$ are reported in \Cref{table:convergence-tvnorm}, along with the results of \cite{Bubeck:2015}.
The dependency of $n$ towards $\CUnfK, \CDeuxfK$ is presented in \Cref{table:convergence-tvnorm-cunkf-cdeuxfk}.
A detailed table is provided in \Cref{appendix:details-table-tv}.

\begin{table}
\begin{center}
\begin{tabular}{|c|c|c|c|c|}
\hline 
Upper bound on $n$ to get $\tvnorm{\delta_{\xstar}  \RKer_{\gaStep}^n-\pi} \leq \varepsilon$  & $d \rightarrow \plusinfty $ & $\varepsilon \rightarrow 0$ & $\rext \rightarrow \plusinfty $ & $\rin \rightarrow 0$ \\ 
\hline 
\Cref{propo:finite-measure-MY} and \Cref{theo:convergence_TV_dec-stepsize-StV} & $\softO(d^5)$ & $\softO(\varepsilon^{-6})$ & $\softO(\rext^4)$ & $\softO(\rin^{-4})$ \\ 
\hline 
\cite[Theorem 1]{Bubeck:2015} $\pi$ uniform on $\convSet$ & $\softO(d^7)$ & $\softO(\varepsilon^{-8})$ & $\softO(\rext^6)$ & $\softO(\rin^{-6})$ \\
\hline
\cite[Theorem 1]{Bubeck:2015} $\pi$ log concave & $\softO(d^{12})$ & $\softO(\varepsilon^{-12})$ & $\softO(\rext^{18})$ &  $\softO(\rin^{-18})$ \\
\hline
\end{tabular} 
\end{center}
\caption{dependency of $n$ on $d$, $\varepsilon$, $\rext$ and $\rin$ to get $\tvnorm{\delta_{\xstar}  \RKer_{\gaStep}^n-\pi} \leq \varepsilon$}
\label{table:convergence-tvnorm}
\end{table}

\begin{table}
\begin{center}
\begin{tabular}{|c|c|c|}
\hline 
Upper bound on $n$ to get $\tvnorm{\delta_{\xstar}  \RKer_{\gaStep}^n-\pi} \leq \varepsilon$  & $\CUnfK \rightarrow 0$ & $\CDeuxfK \rightarrow \plusinfty$ \\ 
\hline 
\Cref{propo:finite-measure-MY} and \Cref{theo:convergence_TV_dec-stepsize-StV} & $\softO(\CUnfK^{-4})$ & $\softO(\CDeuxfK^{4})$ \\
\hline 
\end{tabular} 
\end{center}
\caption{dependency of $n$ on $\CUnfK$ and $\CDeuxfK$ to get $\tvnorm{\delta_{\xstar}  \RKer_{\gaStep}^n-\pi} \leq \varepsilon$}
\label{table:convergence-tvnorm-cunkf-cdeuxfk}
\end{table}

\subsection{Convergence in Wasserstein distance for strongly convex $\fU$}
\label{subsec:convergence-strongly-convex}

In this section, $\fU$ is assumed to satisfy an additional assumption.

\begin{assumption}\label{assumption:fStronglyConvex}
$\fU :\rset^d \mapsto \rset$ is $\m$-strongly convex 
, \ie~there exists $\m >0$ such that for all $x,y\in\rset^d$,
\begin{equation}\label{eq:mfStronglyConvex}
 \fU (y) \geq \fU (x)+  \ps{ \nabla \fU (x)}{y-x} +  (\m/2) \norm[2]{x-y} \eqsp.
\end{equation}
\end{assumption}
Note that under \Cref{assumption:fStronglyConvex}, $\Uml$ defined in \eqref{eq:def-Uml} is $\m$-strongly convex as well.
The following \Cref{prop:wasserstein-epsilon} relies on the convergence analysis in Wasserstein distance done in \cite{durmus:highdimULA2016}, which assumes that $\fU$ is strongly convex. It may be possible to extend the range of validity of these results but this work goes beyond the scope of this paper.

\begin{proposition}
\label{prop:wasserstein-epsilon}
Assume \Cref{assumption:form-potential} and \Cref{assumption:fStronglyConvex}. 
Let $\lambdaMY>0$, $\LUl$ be the Lipschitz constant of $\nabla \Uml$ defined in \eqref{eq:def-Uml} and $\kappa=(2\m\LUl)(\m+\LUl)^{-1}$.
Let $\varepsilon >0$ and $x\in\rset^d$. We have,
\begin{equation*}
W_2(\delta_x \RKer_{\gaStep}^n, \pil) \leq \varepsilon \eqsp ,
\end{equation*}
provided that,
\begin{align*}
\gaStep &\leq \frac{\m}{\LUl^2}\defEns{-\frac{13}{12} + \left[\left(\frac{13}{12}\right)^2+\frac{\varepsilon^2 \kappa^2}{8\m d}\right]^{1/2} } \wedge \frac{1}{\m+\LUl} \eqsp , \\
n &\geq 2(\kappa\gaStep)^{-1}\defEns{-\log(\varepsilon^2/4)+\log \parenthese{\norm[2]{x-\xstar}+d/\m}} \eqsp .
\end{align*}
\end{proposition}

\begin{proof}
The proof is postponed to \Cref{sec:proof-prop-wasserstein-epsilon}.
\end{proof}

Combining \Cref{propo:wasserstein-bounds} and \Cref{prop:wasserstein-epsilon} determines the stepsize $\gamma$ and the number of samples $n$ to get $W_1(\delta_{\xstar}  \RKer_{\gaStep}^n, \pi) \leq \varepsilon$. $\lambdaMY$ is chosen of order $\varepsilon^2 \CUnfK^2 \rin^2 d^{-2} \rext^{-2}$ under \Cref{assumption:fK}-\ref{item:assum-fK-1} and $\varepsilon^2 \rin^2 \rext^{-2} \min(d^{-2}, \CDeuxfK^{-2})$ under \Cref{assumption:fK}-\ref{item:assum-fK-2}.
The orders of magnitude of $n$ in $d,\varepsilon,\rext,\rin, \CUnfK, \CDeuxfK$ are reported in \Cref{table:convergence-wasserstein,table:convergence-wasserstein-cunkf-cdeuxfk}.

\begin{table}
\begin{center}
\begin{tabular}{|c|c|c|c|c|}
\hline 
Upper bound on $n$ to get $W_1(\delta_{\xstar}  \RKer_{\gaStep}^n, \pi) \leq \varepsilon$  & $d \rightarrow \plusinfty $ & $\varepsilon \rightarrow 0$ & $\rext \rightarrow \plusinfty $ & $\rin \rightarrow 0$ \\ 
\hline 
\Cref{propo:wasserstein-bounds}-\ref{item:propo:wasserstein_3} and \Cref{prop:wasserstein-epsilon} & $\softO(d^5)$ & $\softO(\varepsilon^{-6})$ & $\softO(\rext^4)$ & $\softO(\rin^{-4})$ \\
\hline
\end{tabular} 
\end{center}
\caption{dependency of $n$ on $d$, $\varepsilon$, $\rext$ and $\rin$ to get $W_1(\delta_{\xstar}  \RKer_{\gaStep}^n, \pi) \leq \varepsilon$}
\label{table:convergence-wasserstein}
\end{table} 

\begin{table}
\begin{center}
\begin{tabular}{|c|c|c|}
\hline 
Upper bound on $n$ to get $W_1(\delta_{\xstar}  \RKer_{\gaStep}^n, \pi) \leq \varepsilon$  & $\CUnfK \rightarrow 0$ & $\CDeuxfK \rightarrow \plusinfty$ \\ 
\hline 
\Cref{propo:wasserstein-bounds}-\ref{item:propo:wasserstein_3} and \Cref{prop:wasserstein-epsilon} & $\softO(\CUnfK^{-4})$ & $\softO(\CDeuxfK^{4})$ \\
\hline 
\end{tabular} 
\end{center}
\caption{dependency of $n$ on $\CUnfK$ and $\CDeuxfK$ to get $W_1(\delta_{\xstar}  \RKer_{\gaStep}^n, \pi) \leq \varepsilon$}
\label{table:convergence-wasserstein-cunkf-cdeuxfk}
\end{table}


\section{Numerical experiments}
\label{sec:numerical-exp}

In this section we illustrate MYULA with the following three numerical
experiments: computation of the volume of a high-dimensional convex
set, sampling from a truncated multivariate Gaussian distribution, and
Bayesian inference with the constrained LASSO model. We benchmark our
results with model-specific specialised algorithms, namely the
hit-and-run algorithm \cite{lovasz:vempala:2006} for set volume
computation, the wall HMC (WHMC) \cite{pakman2014exact} for truncated
Gaussian models, and the auxiliary-variable Gibbs sampler for the
Bayesian lasso model \cite{park:casella:2008a}. Where relevant we also compare with the Random Walk
Metropolis Hastings (RWM) algorithm.

First we consider the computation of the volume of a high-dimensional hypercube. In a manner akin to \cite{cousins-code}, to apply MYULA  to this problem we use an annealing strategy involving truncated Gaussian distributions whose variance is gradually increased at each step $i\in\nset$ of the annealing process. Precisely, for $M\in \nset^\star$ and $i\in\{0,\ldots,M-1\}$, the potential $U_i$ \eqref{eq:definition-U} of the phase $i$ is given for all $x\in\rset^d$ by, $U_i(x)=(2\sigma_i^2)^{-1} \norm{x}^2 + \ind_{\convSet}$ where $\convSet=\ccint{-1,1}^d$. Observing that,
\begin{equation}\label{eq:volume-1}
\frac{\int_{\rset^d} \rme^{-U_{i+1}(x)} \rmd x}{\int_{\rset^d} \rme^{-U_{i}(x)} \rmd x} = \pi_i \parenthese{g_i} \eqsp , \quad
g_i(x) = \rme^{2^{-1}(\sigma_i^{-2}-\sigma_{i+1}^{-2})\norm{x}^2} \eqsp , 
\end{equation}
where $\pi_i$ is the probability measure associated with $U_i$, the volume of $\convSet$ is
\begin{equation*}
\vol(\convSet) = \prod_{i=0}^{M-1} \pi_i (g_i) \int_{\rset^d} \rme^{-U_{0}(x)} \eqsp ,
\end{equation*}
where $U_M=\ind_\convSet$. To use MYULA we consider for all $i\in\{0,\ldots,M-1\}$ the potential $U_i^{\lambda_i}$ defined for all $x\in\rset^d$ by $U_i^{\lambda_i}(x)=(2\sigma_i^2)^{-1} \norm{x}^2 + \ind^{\lambda_i}_{\convSet}$ where $\ind^{\lambda_i}_{\convSet}$ is given by \eqref{eq:def-indKMY}. We choose the step-size $\gamma_i$ proportional to $1/\{d\max(d,\sigma_i^{-1})\}$ and the regularization parameter $\lambda_i$ is set equal to $ 2 \gamma_i$. The counterpart of \eqref{eq:volume-1} is then 
\begin{equation*}
\frac{\int_{\rset^d} \rme^{-U_{i+1}^{\lambda_{i+1}}(x)} \rmd x}{\int_{\rset^d} \rme^{-U_i^{\lambda_{i}}(x)} \rmd x} = \pi_{i}^{\lambda_i} \parenthese{g_i^{\lambda_i}} \eqsp , \quad
g_i^{\lambda_i}(x) = \rme^{2^{-1}(\sigma_i^{-2}-\sigma_{i+1}^{-2})\norm{x}^2+\ind^{\lambda_i}_{\convSet}-\ind^{\lambda_{i+1}}_{\convSet}} \eqsp , 
\end{equation*}
where $\pi_i^{\lambda_i}$ is the probability measure associated with $U_i^{\lambda_i}$, and the volume of $\convSet$ is
\begin{equation*}
\vol(\convSet) = \prod_{i=0}^{M-1} \pi_i^{\lambda_i} (g_i^{\lambda_i}) \int_{\rset^d} \rme^{-U_{0}^{\lambda_0}(x)} \eqsp ,
\end{equation*}
where $U_M^{\lambda_M}=U_M=\ind_\convSet$.

\begin{figure}
\begin{center}
\includegraphics[scale=0.7]{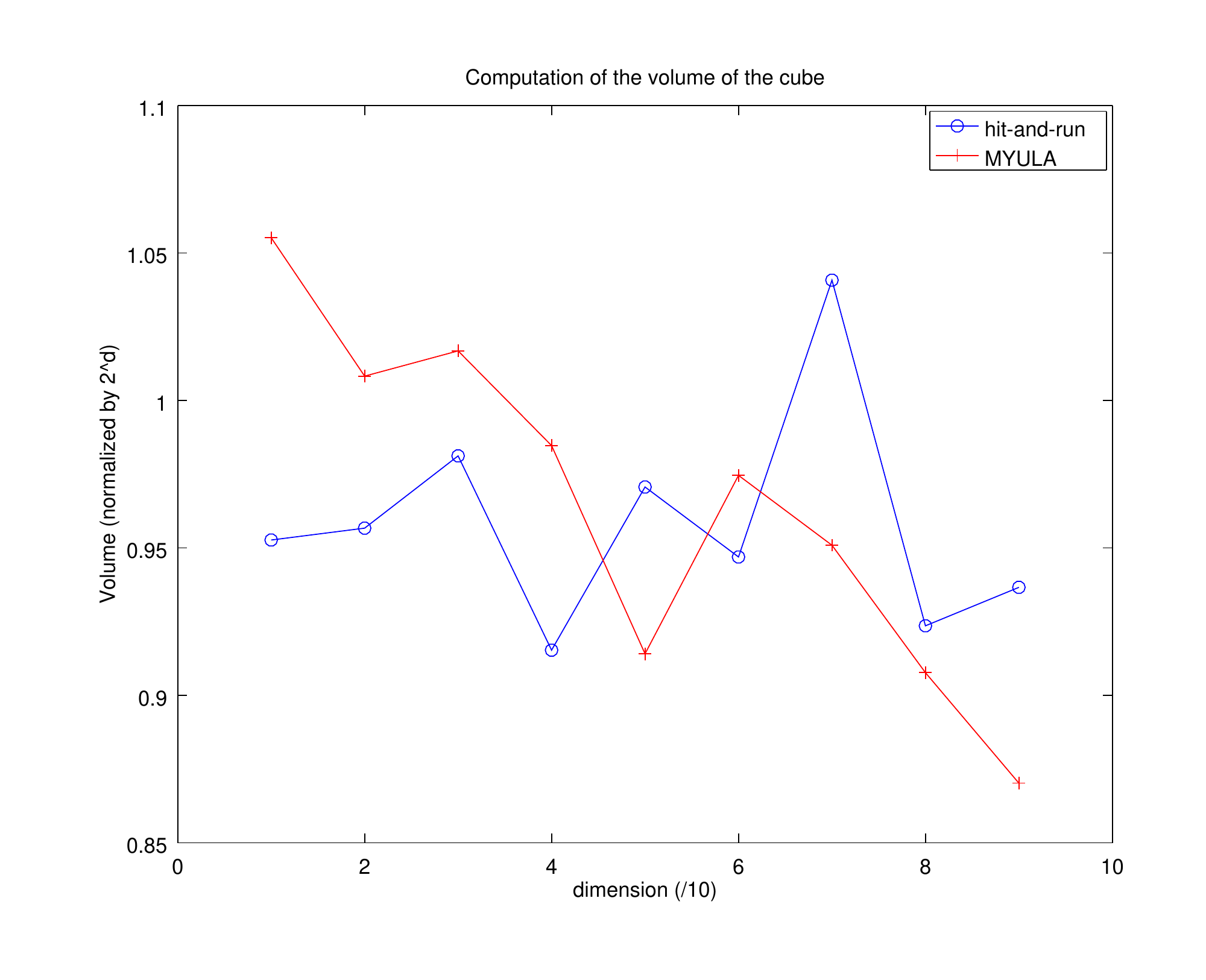} 
\end{center}
\caption{\label{figure:volume-convex-2} Computation of the volume of the cube with MYULA and hit-and-run algorithm.}
\end{figure}

Figure \ref{figure:volume-convex-2} shows the volume estimates (over 10 experiments)
obtained with MYULA and the hit-and-run algorithm for a unit
hypercube of dimension $d$ ranging from $d=10$ to $d = 90$ (to
simplify visual comparison the estimates are normalised w.r.t. the
true volume). Observe that the estimates of MYULA are in agreement
with the results of the hit-and-run algorithm, which serves as a 
benchmark for this problem. The outputs of both algorithms are at similar distances with respect to the true value $1$.

Moreover, the second experiment we consider is the simulation from a $d$-dimensional truncated Gaussian distribution restricted on a convex set $\convSet_d$, with mode zero at the boundary of the set, and covariance matrix $\Sigma$ with $(i,j)$th element given by $(\Sigma)_{i,j}=1/(1+\absolute{i-j})$. Let $\bfbeta\in\rset^d$. The potential $U$, given by \eqref{eq:definition-U} and associated with the density $\pi(\bfbeta)$, is given by $U(\betabf) = (1/2)\ps{\betabf}{\Sigma^{-1}\betabf} + \ind_{\convSet_d }(\betabf)$. We consider three scenarios of increasing dimension: $d = 2$ with $\convSet_2 = \ccint{0,5} \times \ccint{0,1}$, $d=10$ with $\convSet_{10} = \ccint{0,5} \times \ccint{0,0.5}^{9}$, and $d=100$ with $\convSet_{100} = \ccint{0,5} \times \ccint{0,0.5}^{99}$. We generate $10^6$ samples for MYULA, $10^5$ samples for WHMC, and $10^6$  samples for RWM (in all cases the initial $10\%$ is discarded as burn-in period). Regarding algorithm parameters, we set $\gamma = 1/1000$ and $\lambda = 2 \gamma$ for MYULA, and adjust the parameters of RWM and WHMC such that their acceptance rates are approximately $25\%$ and $70\%$. 

\begin{figure}
\begin{center}
\includegraphics[scale=0.6]{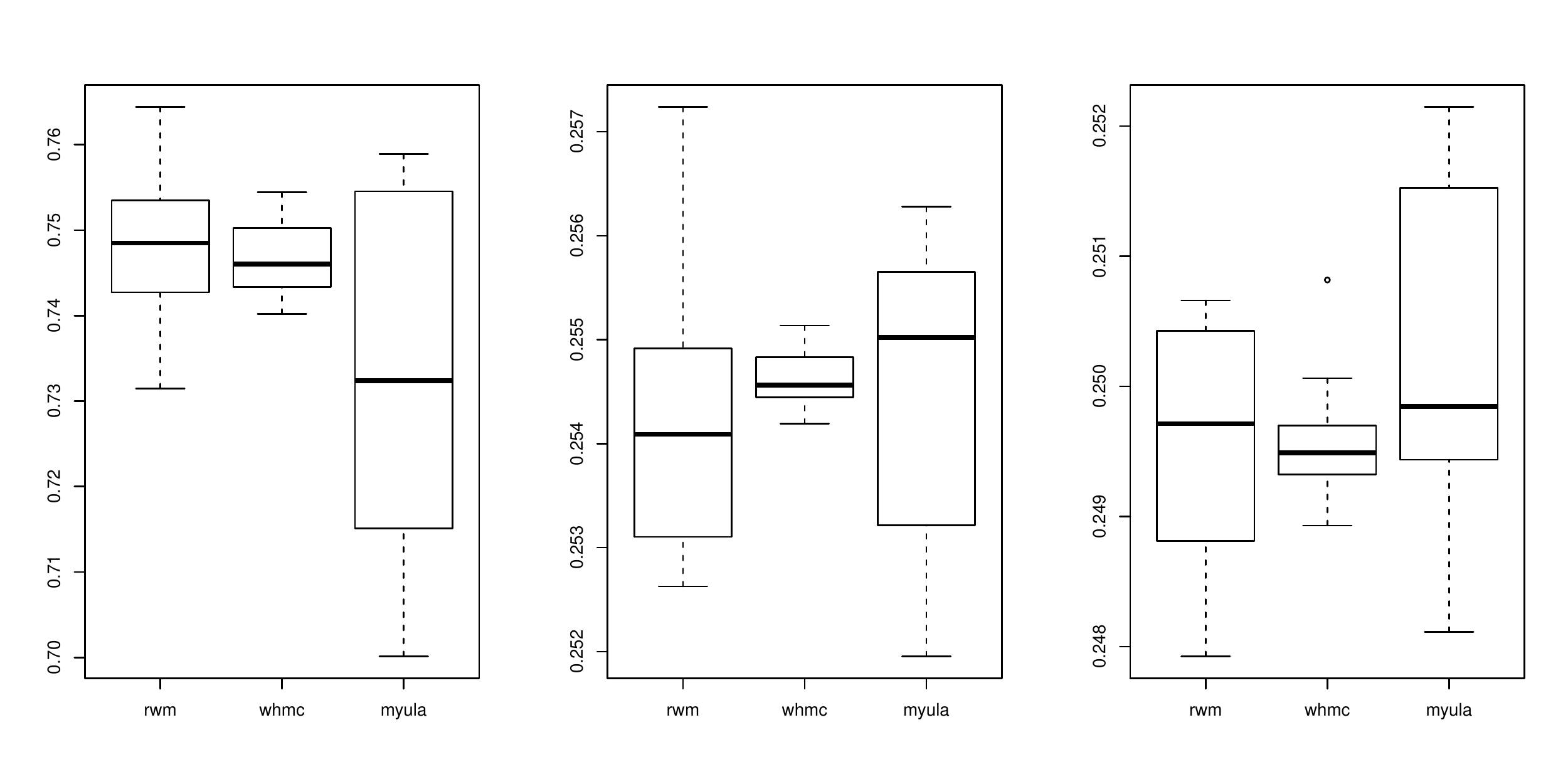} 
\end{center}
\caption{\label{figure:TMGD10} Boxplots of $\beta_1,\beta_2,\beta_3$ for the truncated Gaussian variable in dimension 10.}
\end{figure}

\begin{figure}
\begin{center}
\includegraphics[scale=0.6]{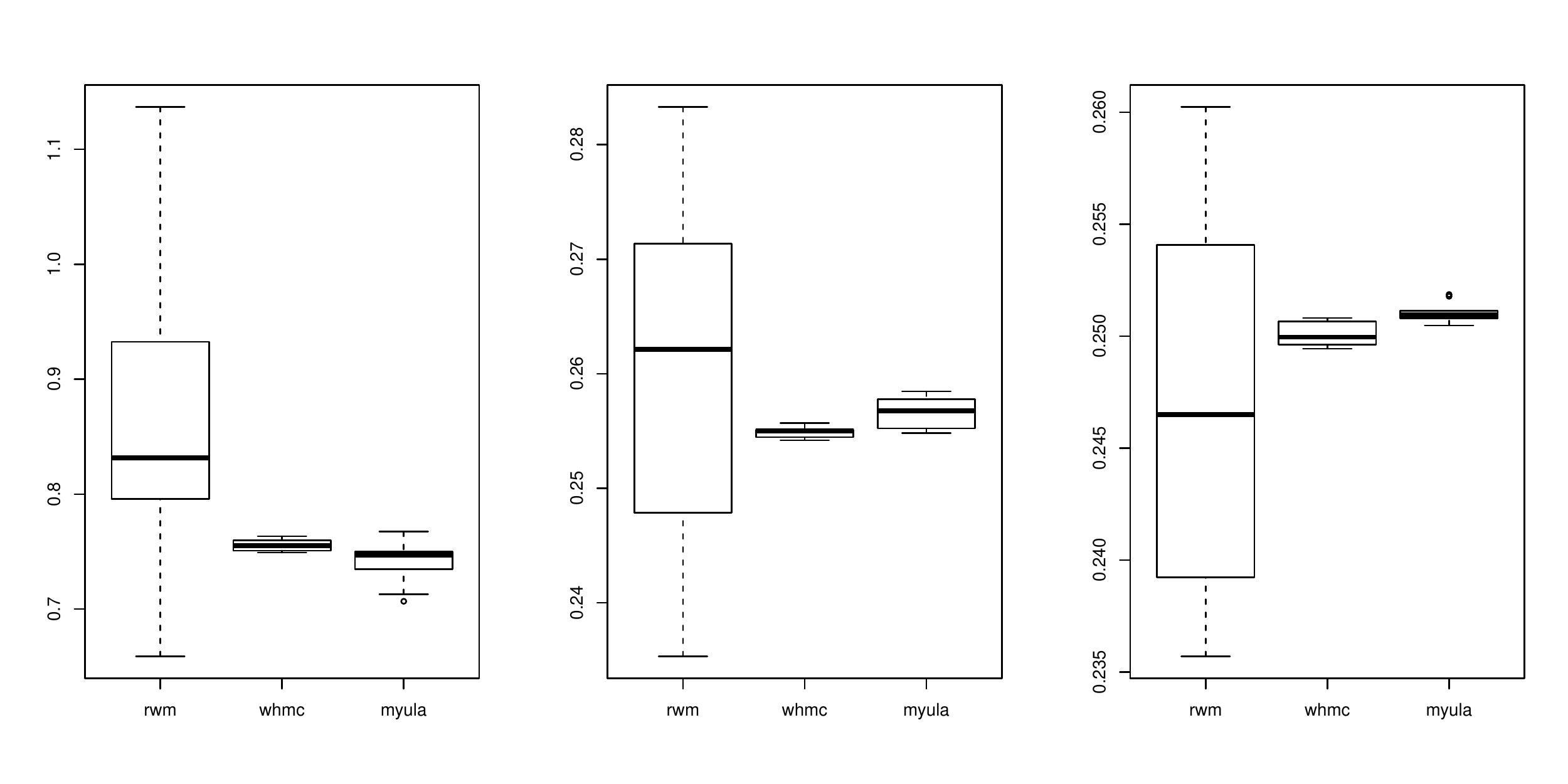} 
\end{center}
\caption{\label{figure:TMGD100} Boxplots of $\beta_1,\beta_2,\beta_3$ for the truncated Gaussian variable in dimension 100.}
\end{figure}

\Cref{TMG-D2} shows the results obtained with each method for the model $d = 2$, and by performing $100$ repetitions to obtain $95\%$ confidence intervals. For this model we also report a solution by a cubature integration \cite{package-cubature} which provides a ground truth. Moreover, \Cref{figure:TMGD10} and \Cref{figure:TMGD100} show the results for the first three coordinates of $\bfbeta$ (i.e., $\beta_1,\beta_2,\beta_3$) for $d=10$ and $d=100$ respectively. Observe the good performance of MYULA  as dimensionality increases, particularly in the challenging case $d=100$ where it performs comparably to the specialised algorithm WHMC.

\begin{table}[ht]
\centering
\begin{tabular}{l|c|c}
  \hline
Method & Mean & Covariance \\ 
\hline
Truth &
$\begin{bmatrix}
0.790 \\
0.488
\end{bmatrix}$ &
$\begin{bmatrix}
0.326 & 0.017 \\
0.017 & 0.080
\end{bmatrix}$ \\
  \hline
RWM & 
$\begin{bmatrix}
0.791 \pm 0.013 \\
0.486 \pm 0.002
\end{bmatrix}$
&
$\begin{bmatrix}
0.330 \pm 0.011 &
0.017 \pm 0.002 \\
0.017 \pm 0.002 &
0.080 \pm 0.0003
\end{bmatrix}$ \\
\hline
WHMC &
$\begin{bmatrix}
0.789\pm0.005 \\
0.490\pm0.005
\end{bmatrix}$
&
$\begin{bmatrix}
0.324\pm0.008 &
0.017\pm0.002 \\
0.017\pm0.002 &
0.079\pm0.0007 
\end{bmatrix}$ \\
\hline
MYULA &
$\begin{bmatrix}
0.758\pm0.052 \\
0.484\pm0.016
\end{bmatrix}$ &
$\begin{bmatrix}
0.309\pm0.038 &
0.017\pm0.009 \\
0.017\pm0.009 &
0.088\pm0.002
\end{bmatrix}$ \\
\hline
\end{tabular}
\caption{Mean and covariance of $\bfbeta$ in dimension 2 obtained by RWM, WHMC and MYULA.} 
\label{TMG-D2}
\end{table} 

Finally, we also report an experiment involving the analysis of a real dataset with an $\ell_1$-norm constrained Bayesian LASSO model (\ie~ least squares regression subject to  an $\ell_1$-ball constraint). Precisely, the
observations $Y = \defEns{Y_1, \ldots, Y_n} \in \rset^n$, for $n \geq
1$, are assumed to be distributed from the Gaussian distribution with
mean $X \betabf$ and covariance matrix $\sigma^2 \operatorname{I}_n$,
where $X \in \rset^{n \times d}$ is the design matrix, $\betabf \in
\rset^d$ is the regression parameter, $\sigma^2 >0$ and
$\operatorname{I}_n$ is the identity matrix of dimension
$n$. The prior on $\betabf$ is the uniform distribution over the $\ell_1$
ball, $\ballO{0}{\seuil} = \{ \betabf\in\rset^d \, \; \,
\norm{\betabf}_1\leq \seuil \}$, for $\seuil >0$, where
$\norm{\betabf}_1 = \sum_{i=1}^d \abs{\betabf_i}$, $\betabf_i$ is the
$i$-th component of $\betabf$.  The potential $U^{\seuil}$, for
$\seuil >0$, associated with the posterior distribution is given for
all $\betabf\in\rset^d$ by $U^{\seuil}(\betabf) = \normLigne[2]{Y-X
  \betabf} + \ind_{\ballO{0}{s}}(\betabf)$. We consider in our experiment the
diabetes data set\footnote{http://archive.ics.uci.edu/ml/datasets/Pima+Indians+Diabetes}, which consists
in $n=442$ observations and $d= 10$ explanatory variables.

\begin{figure}
\begin{center}
\includegraphics[scale=0.6]{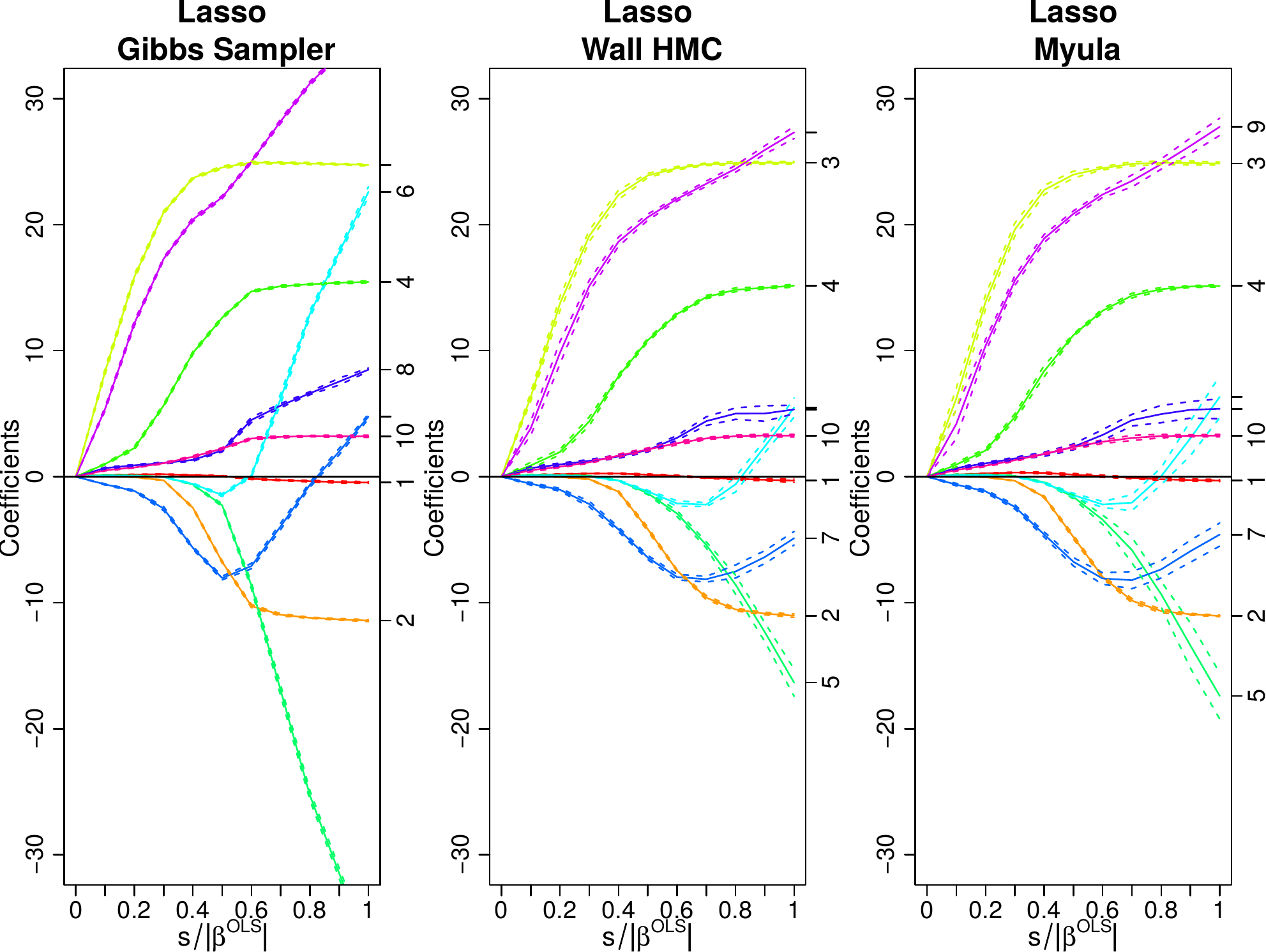} 
\end{center}
\caption{\label{figure:lasso} Lasso path for the Gibbs sampler, Wall HMC and MYULA algorithms.}
\end{figure}

\Cref{figure:lasso} shows the ``LASSO paths" obtained using MYULA,
the WHMC algorithm, and with the specialised Gibbs sampler of \cite{park:casella:2008a} (these paths are the posterior marginal medians associated with
$\pi^{\seuil}$ for $\seuil = t\norm{\bfbetaOLS}_1$, $t\in\ccint{0,1}$,
and where $\bfbetaOLS$ is the estimate obtained by the ordinary least square
regression). The dot lines represent the confidence
interval at level $95\%$, obtained by performing $100$ repetitions. MYULA estimates were obtained by using $10^5$ samples (with the initial $10^4$ samples discarded as burn-in period) and stepsize $\seuil^{3/2} \times 10^{-5}$. WHMC estimates were obtained by using $10^4$ samples (with the initial $10^3$ samples discarded as burn-in period), and by adjusting parameters to achieve an acceptance rate of approximately $90\%$. Finally, the Gibbs sampler is targeting an unconstrained LASSO model with prior $\betabf \mapsto
(2\seuil)^{-d} \rme^{-\norm{\betabf}_1/\seuil}$, for $\seuil >0$.


\section{Proofs}
\label{sec:proofs}

\subsection{Proof of \Cref{lemma:pilambda-defined}}
\label{sec:proof-lemma-pilambda-defined}

Since $\fU$ is a (proper) convex function, there exist $a\in\rset$, $b\in\rset^d$ such that $\fU(x) \geq a + \crochet{b, x}$ \cite[Theorem 23.4]{rockafellar2015convex}. By \Cref{assumption:K} and a straightforward calculation, for $\norm{x} \geq \rext + 4\lambda\norm{b} +2\defEns{\lambda(\absolute{a}+\rext\norm{b})}^{1/2}$, we have,
\[
 \Uml(x) \geq (4\lambda)^{-1} (\norm{x} - \rext)^2 \eqsp,
\]
which concludes the proof.

\subsection{Proof of \Cref{lemma:KTilde}}
\label{sec:proof-lemmaKTilde}

Under \Cref{assumption:K}, $0\in\convSetTilde$. Let $x_1,x_2 \in\convSetTilde$ and $t\in\ccint{0,1}$. We have by definition of $\convSetTilde$ \eqref{eq:def-convSetTilde} that $\ball{t x_1 + (1-t) x_2}{\rin} \subset t\ball{x_1}{\rin} + (1-t) \ball{x_2}{\rin} \subset \convSet$, which implies that $\convSetTilde$ is convex. 

To show that $\convSetTilde$ is close, it is enough to show that
$\convSetTilde = \defEns{x\in\convSet \eqsp | \eqsp \text{dist}(x,
  \convSet^c) \geq \rin}$ where $\text{dist}(x,\convSet^c) =
\inf_{y\in\convSet^c} \norm{x-y}$ since $x \mapsto
\text{dist}(x,\convSet^c)$ is Lipschitz continuous. First by definition, we have
$\convSetTilde \subset \defEns{x\in\convSet \eqsp | \eqsp
  \text{dist}(x, \convSet^c) \geq \rin}$. To show the converse, let
$x\in\defEns{y\in\convSet \eqsp | \eqsp \text{dist}(y, \convSet^c)
  \geq \rin}$. Then, $\ballO{x}{\rin} \subset \convSet$, where
$\ballO{x}{\rin} = \defEns{y\in\rset^d \eqsp | \eqsp \norm{y-x} <
  \rin}$, which yields $\ball{x}{\rin} \subset \convSet$ since $\convSet$ is assumed to be close. This result then concludes the proof by definition of $\convSetTilde$. 

\subsection{Proof of \Cref{propo:finite-measure-MY}}
\label{sec:proof-pi-pilambda-tv}

\begin{enumerate}[label=\alph*), wide=0pt, labelindent=\parindent]
\item 
  By a direct calculation, we have:
  \begin{align}
\label{eq:tv-1}
\tvnorm{\piml-\pi} = \int_{\rset^d} \abs{\pi(x)-\piml(x)}\rmd x  &= 2 \parenthese{1+\defEns{\int_{\convSet^c} \rme^{-\Uml(x)} \rmd x}^{-1}\int_{\convSet}\rme^{-\fU(x)} \rmd x}^{-1} \\
\label{eq:indi_conv_proof1}
&\leq 2 \parenthese{1+\exp \parenthese{\min_{\convSet^{c}}(f)-\max_{\convSet}(f)} A }^{-1} \eqsp.
  \end{align}
  where 
  \begin{equation}
  A = \left. \vol(\convSet) \Big/ \int_{\convSet^c} \rme^{-(2  \lambdaMY)^{-1} \norm[2]{x- \proj{x}{\convSet}}} \rmd x \right. \eqsp .
  \end{equation}
The conclusion follows then from \eqref{eq:Wills} and \Cref{assumption:fK}-\ref{item:assum-fK-1}.

\item 
\label{item:proof-TV-2}

We give two proofs for this result, which both consist in lower bounding $A$.
The obtained bounds are identical up to an universal constant.
The first one is simpler and was suggested by a referee. The second one is more involved ; however, it has the benefit of establishing the relation between the intrinsic volumes of $\convSet$ and the bound on the total variation norm.

Under \Cref{assumption:K}, we have $\convSet + \ball{0}{t} \subset (1+t/\rin) \convSet$ and using \eqref{eq:kampf},
\begin{align*}
\int_{\convSet^c} \rme^{-(1/2\lambda)\norm{x-\projK x}^2} \rmd x &\leq 
\defEns{\int_{\rset_{+}} \vol(\convSet(1+t/\rin))\lambda^{-1}t\rme^{-t^2/(2\lambda)} \rmd t - \vol(\convSet) } \\
&= \vol(\convSet) \defEns{\int_{\rset_{+}} (1+t/\rin)^d\lambda^{-1}t\rme^{-t^2/(2\lambda)} \rmd t - 1 } \\
&= \vol(\convSet) \sum_{i=1}^d \binom{d}{i} \parenthese{\frac{\sqrt{2\lambda}}{\rin}}^i \Gammabf(1+i/2) \\
&\leq \vol(\convSet) \sum_{i=1}^d \parenthese{\frac{\sqrt{2\lambda}d}{\rin}}^i \eqsp ,
\end{align*}
where the second equality follows from developping $(1+t/\rin)^d$, making the change of variable $t \mapsto t^2/(2\lambda)$ and using the Gamma function and the last inequality from $\binom{d}{i} \Gammabf(1+i/2) \leq d^i$ for $i\in\defEns{1,\ldots,d}$.
For $\lambda\in\ocint{0,\rin^2 d^{-2} / 8}$, we get
\begin{equation*}
A^{-1} \leq 
\sum_{i=1}^d \parenthese{\frac{\sqrt{2\lambda}d}{\rin}}^i
\leq \frac{2\sqrt{2\lambda}d}{\rin} \eqsp .
\end{equation*}
Combining it with \eqref{eq:indi_conv_proof1} and \Cref{assumption:fK}-\ref{item:assum-fK-1} concludes the proof.

For the second proof, it is necessary to introduce first a generalized notion of the intrinsic volumes \eqref{eq:Steiner_formula}, the mixed volumes. Let $\ensConvSet$ be the class of convex bodies of $\rset^d$, $\convSet_1, \ldots, \convSet_m \in \ensConvSet$ and $\lambda_1, \ldots, \lambda_m \geq 0$. 
By \cite[Theorem 5.1.7]{schneider:2013}, there is a nonnegative symmetric function $\VolMix:(\ensConvSet)^d \to \rset_+$, the mixed volume, such that,
\begin{equation}\label{eq:def-mixed-volumes}
\vol(\lambda_1 \convSet_1 + \ldots + \lambda_m \convSet_m) = \sum_{i_1, \ldots, i_d = 1}^m \lambda_{i_1}\ldots\lambda_{i_d}\VolMix(\convSet_{i_1}, \ldots, \convSet_{i_d}) \eqsp .
\end{equation}
Let $m>1$, $a_1, \ldots, a_m \geq 0$ and $\convSet_1, \ldots, \convSet_m, \mathsf{L}$ be $(m+1)$ convex bodies in $\rset^d$ such that $\convSet_1 \subset \mathsf{L}$. By unicity of the coefficients of the polynomial in $\lambda_1,\ldots, \lambda_m$ \eqref{eq:def-mixed-volumes} and \cite[p.282]{schneider:2013}, we have:
\begin{align}
\label{eq:posHomogeneity}
\VolMix(a_1 \convSet_1, \ldots, a_m \convSet_m) &= \parenthese{\prod_{i=1}^m a_i} \VolMix(\convSet_1, \ldots, \convSet_m) \eqsp , \\
\label{eq:monotonicity}
\VolMix(\convSet_1, \convSet_2, \ldots, \convSet_m) &\leq \VolMix(\mathsf{L},\convSet_2, \ldots, \convSet_m) \eqsp .
\end{align}
Denote by $\ballUn$ the unity ball of $\rset^d$, $\ballUn=\ball{0}{1}$. Taking $m=2, \convSet_1=\convSet, \convSet_2=\ballUn, \lambda_1 = 1, \lambda_2 = t$ in \eqref{eq:def-mixed-volumes}, we get:
\begin{equation}\label{eq:def-quermass-mixVol}
\vol(\convSet + \ball{0}{t}) = \sum_{i=0}^{d} t^{i} \binom{d}{i} \VolMix(\convSet[d-i], \ballUn[i]) \eqsp,
\end{equation}
where for a set $A\subset \rset^d$, the notation $A[i]$ means $A$ repeated $i$ times: $A[i]=A, \ldots, A$ $i$ times.
The quermassintegrals of $\convSet$ are defined for $i\in\defEns{0,\ldots,d}$ by $\Querm_i(\convSet)=\VolMix(\convSet[d-i], \ballUn[i])$ \cite[equation 5.31]{schneider:2013}. We get then by \eqref{eq:def-quermass-mixVol} and \eqref{eq:Steiner_formula},
\begin{equation}\label{eq:defQuermass}
\binom{d}{i} \Querm_i(\convSet) = \volball_i \VolInt_{d-i}(\convSet) \eqsp,
\end{equation}
where $\volball_i$ is given by \eqref{eq:vol_ball}.

The proof consists then in identifying an upper bound on $\VolInt_i(\convSet)(\vol \convSet)^{-1}$ for $i\in\defEns{0,\ldots,d}$. 
First, the sequence $\{i! \VolInt_i(\convSet)\}_{0 \leq i \leq d}$ is shown to be log-concave, \ie~for $i\in\defEns{1,\ldots,d-1}$
\begin{equation}\label{eq:logconcV}
\left(i!\VolInt_i(\convSet)\right)^2 \geq (i+1)!\VolInt_{i+1}(\convSet) (i-1)!\VolInt_{i-1}(\convSet) \eqsp .
\end{equation}
The Aleksandrov-Fenchel inequality \cite[equation 7.66]{schneider:2013} states, for $i\in\defEns{1,\ldots,d-1}$,
\begin{equation}\label{eq:AlexFenchW}
\Querm_i(\convSet)^2 \geq \Querm_{i-1}(\convSet)\Querm_{i+1}(\convSet) \eqsp .
\end{equation}
By \eqref{eq:vol_ball}, $\volball_i/\volball_{i-2} = (2\uppi)/i$ and the log convexity of the gamma function, we get for $i\in\defEns{1,\ldots,d-1}$:
\begin{equation}\label{eq:AlexFenchCube}
\frac{1}{i+1}\frac{\volball_i}{\volball_{i+1}}=\frac{1}{i}\frac{\volball_{i-2}}{\volball_{i-1}} \leq \frac{1}{i}\frac{\volball_{i-1}}{\volball_i} \eqsp .
\end{equation}
Combining \eqref{eq:AlexFenchCube}, \eqref{eq:AlexFenchW} and \eqref{eq:defQuermass} shows \eqref{eq:logconcV}.

The log-concavity of $\{i! \VolInt_i(\convSet)\}_{0 \leq i \leq d}$ gives for $i\in\defEns{0,\ldots,d-1}$,
\begin{equation}\label{eq:InequalityVolInt}
\frac{\VolInt_i(\convSet)}{\VolInt_{i+1}(\convSet)} \leq \frac{\VolInt_{d-1}(\convSet)}{\vol(\convSet)} = \frac{d}{2}\frac{\Querm_1(\convSet)}{\Querm_0(\convSet)} \eqsp .
\end{equation}
Combining the definition of the quermassintegrals, \eqref{eq:posHomogeneity}, \eqref{eq:monotonicity} and \Cref{assumption:K} give:
\begin{equation}\label{eq:MonoQuerm}
\rin \Querm_1(\convSet) = \VolMix(\convSet,\ldots,\convSet,\ball{0}{\rin}) \leq \VolMix(\convSet,\ldots,\convSet,\convSet) = \Querm_0(\convSet) \eqsp .
\end{equation}
By \eqref{eq:MonoQuerm} and \eqref{eq:InequalityVolInt}, we get:
\begin{equation}
\label{eq:boundDconst}
\DconstConvInt(\convSet,\lambdaMY) \leq \sum_{i=1}^d \defEns{d\rin^{-1} (\uppi\lambdaMY /2)^{1/2}}^i \eqsp ,
\end{equation}
where $\DconstConvInt(\convSet,\lambdaMY)$ is defined in \eqref{eq:def-DKlambda}.
For all $\lambdaMY \in\ooint{0,2\uppi^{-1}(\rin /d)^2}$, \eqref{eq:propo:dist_TV_MY_1} gives then,
\begin{equation*}
\tvnorm{\piml-\pi} \leq 2 \defEns{1+\exp \parenthese{\min_{\convSet^{c}}(f)-\max_{\convSet}(f)}\parenthese{\defEns{d\rin^{-1} (\uppi\lambdaMY/2)^{1/2}}^{-1} -1}}^{-1} \eqsp.
\end{equation*}
Using that for all $a,b \in \rset_+^*$, $b \geq 2$, $(1+a(b-1))^{-1} \leq b^{-1}/(b^{-1}+a/2)$ and \Cref{assumption:fK}-\ref{item:assum-fK-1}, we get for $\lambdaMY \in\ooint{0,2\uppi^{-1}(\rin /d)^2}$
\begin{equation*}
\tvnorm{\piml-\pi} \leq  2^{3/2}  (\uppi\lambda)^{1/2}d\rin^{-1} \defEns{(2\uppi\lambda)^{1/2}d\rin^{-1}+ \CUnfK }^{-1} \eqsp .
\end{equation*}

\item
\label{item:proof-TV-3}
The proof consists in using \eqref{eq:tv-1} to bound $\tvnorm{\piml - \pi}$.
In the first step we give an upper bound on $\int_{\rset^d} \rme^{-\Uml(x)} \rmd x / \int_{\convSet} \rme^{-f(x)} \rmd x$.
By Fubini's theorem, similarly to \eqref{eq:kampf} we have
\begin{equation}\label{eq:proof-TV-fK-1}
\int_{\rset^d} \rme^{-\Uml(x)} \rmd x \leq \int_{\rset_+} \int_{\convSet+\ball{0}{t}} \rme^{-f(x)} \lambda^{-1} t\rme^{-t^2/(2 \lambdaMY)} \rmd x \rmd t \eqsp.
\end{equation}
Let $t\geq 0$. By definition of $\convSetTilde$, using \Cref{lemma:KTilde} and $\convSet - \projxK + \ball{0}{t} \subset (1+t/\rin)(\convSet-\projxK)$, we have
\begin{align}
\nonumber
\int_{\convSet+\ball{0}{t}} \rme^{-f(x)} \rmd x 
&= \int_{\convSet - \projxK +\ball{0}{t}} \rme^{-f(x+\projxK)} \rmd x \\
\nonumber
&\leq \int_{(1+t/\rin)(\convSet - \projxK)} \rme^{-f(x+\projxK)} \rmd x \\
\label{eq:proof-TV-fK-2}
&= (1+t/\rin)^d \int_{\convSet - \projxK} \rme^{-f((1+t/\rin)x+\projxK)} \rmd x \eqsp.
\end{align}
By \Cref{assumption:form-potential}-\ref{item:assum-fU-convex} $f$ is convex and therefore for all $x\in\convSet -\projxK$,
\begin{align*}
f((1+t/\rin)x+\projxK) 
&\geq (t/\rin)\defEns{f(x+\projxK)-f(\projxK)}+f(x+\projxK) \\
&\geq  -(\CDeuxfK t)/\rin + f(x+\projxK) \eqsp.
\end{align*}
Combining it with \eqref{eq:proof-TV-fK-1} and \eqref{eq:proof-TV-fK-2}, we get
\begin{equation}\label{eq:tv-2}
\int_{\rset^d} \rme^{-\Uml(x)} \rmd x \leq \parenthese{\int_{\convSet} \rme^{-f(x)} \rmd x}
\int_{\rset_+} (1+t/\rin)^d \rme^{(\CDeuxfK t)/\rin} \lambda^{-1} t \rme^{-t^2/(2 \lambdaMY)} \rmd t \eqsp.
\end{equation}
We now bound $B = \int_{\convSet^c} \rme^{-\Uml(x)} \rmd x / \int_{\convSet} \rme^{-f(x)} \rmd x$.
Using \eqref{eq:tv-2} and an integration by parts, we have
\begin{align*}
B &\leq \int_{\rset_+} \defEns{(1+t/\rin)^d \rme^{(\CDeuxfK t)/\rin} -1} \lambda^{-1} t \rme^{-t^2/(2 \lambdaMY)} \rmd t \\
&\leq \int_{\rset_+} (1+t/\rin)^{d-1} \rme^{(\CDeuxfK t)/\rin} \rin^{-1} \parenthese{d+\CDeuxfK + (\CDeuxfK t)/\rin} \rme^{-t^2/(2 \lambdaMY)} \rmd t \eqsp.
\end{align*}
Since for all $t \geq 0$, $(\CDeuxfK t)/\rin - t^2/(2\lambda) \leq -t^2/(4\lambda) + 4\lambda(\CDeuxfK/\rin)^2$, it holds
\begin{equation*}
B \leq \frac{1}{\rin}\exp\parenthese{4\lambda\parenthese{\frac{\CDeuxfK}{\rin}}^2}
\int_{\rset_+} (1+t/\rin)^{d-1} \parenthese{d+\CDeuxfK + (\CDeuxfK t)/\rin} \rme^{-t^2/(4 \lambdaMY)} \rmd t \eqsp.
\end{equation*}
By developping $(1+t/\rin)^{d-1}$, using the change of variable $t \mapsto t^2/(4\lambda)$ and the definition of the Gamma function, we have
\begin{equation*}
B \leq \frac{2\lambda}{\rin}\exp\parenthese{4\lambda\parenthese{\frac{\CDeuxfK}{\rin}}^2}
\sum_{i=0}^{d-1} \binom{d-1}{i}\parenthese{\frac{2\sqrt{\lambda}}{\rin}}^i 
\defEns{\frac{d+\CDeuxfK}{2\sqrt{\lambda}}\Gammabf\parenthese{\frac{1+i}{2}} + \frac{\CDeuxfK}{\rin}\Gammabf\parenthese{1+\frac{i}{2}}} \eqsp.
\end{equation*}
Using that for all $i\in\defEns{0,\ldots,d-1}$, $\binom{d-1}{i} \Gammabf(1+i/2) \leq d^i$, we get for $\lambda\in\ocint{0, 16^{-1}\rin^2 d^{-2}}$
\begin{equation*}
B \leq \frac{2}{\rin}\exp\parenthese{4\lambda\parenthese{\frac{\CDeuxfK}{\rin}}^2}
\defEns{\sqrt{\lambda}(d+\CDeuxfK) + \frac{2\lambda\CDeuxfK}{\rin} } \eqsp,
\end{equation*}
which combined with \eqref{eq:tv-1} concludes the proof.
\end{enumerate}

\subsection{Proof of \Cref{propo:wasserstein-bounds}}
\label{sec:proof-pi-pilambda-wasserstein}

\begin{enumerate}[label=\alph*), wide=0pt, labelindent=\parindent]
\item 
The proof relies on a control of the Wasserstein distance by a weighted total variation. The arguments are similar to those of \Cref{propo:finite-measure-MY}. \cite[Theorem 6.15]{VillaniTransport} implies:
\begin{equation}\label{eq:vt-wasserstein}
W_1(\pi,\pil) \leq \int_{\rset^d} \norm{x} \absLigne{\pi(x)-\pil(x)}\rmd x = C+D \eqsp,
\end{equation}
where
\begin{equation}\label{eq:def-A-B}
C = \int_{\convSet^{c}} \norm{x} \pil(x) \rmd x \eqsp , \quad 
D = \defEns{1-\frac{\int_\convSet \rme^{-f}}{\int_{\rset^d} \rme^{-\Uml}}} \int_\convSet \norm{x} \pi(x) \rmd x \eqsp .
\end{equation}
We bound these two terms separately.
First using the same decomposition as in \eqref{eq:kampf},  $\norm{x} \leq \rext + \norm{x-\projK x}$ and that for all $t \in \rset_+$, $\convSet + \ball{0}{t} = \{ x \in \rset^d \, : \, \norm{x-\projK x} \leq t\}$, we get 
\begin{align}
\label{eq:WT-1}
  C &=\parenthese{\int_{\rset^d} \rme^{-\Uml}}^{-1} \int_{0}^{\plusinfty} \int_{\convSet^{c}} \rme^{-\fU(x)} \norm{x} t\lambda^{-1}\rme^{-t^2/(2\lambda)} \1_{ \coint{\norm{x-\projK x},\plusinfty}}(t) \:  \rmd x \, \rmd t \\
\label{eq:B-bound-1}
& \leq  \rme^{\max_{\convSet}(f)-\min_{\convSet^{c}}(f)} \int_{0}^{\plusinfty} (\rext+t) t\lambda^{-1} \rme^{-t^2/(2\lambda)} \parenthese{\frac{\vol(\convSet + \ball{0}{t}) - \vol(\convSet)}{\vol(\convSet)}} \rmd t \eqsp.
\end{align}
Combining \eqref{eq:Steiner_formula}-\eqref{eq:B-bound-1}, \Cref{assumption:fK}-\ref{item:assum-fK-1} and using $\VolInt_{d}(\convSet) = \vol(\convSet)$ give
\begin{equation}\label{eq:B-bound-1-2}
C \leq \CUnfK^{-1} \sum_{i=0}^{d-1} \volball_{d-i} \frac{\VolInt_{i}(\convSet)}{\vol(\convSet)} 
\int_{0}^{\plusinfty} (\rext t^{d-i+1}+t^{d-i+2}) \lambda^{-1} \rme^{-t^2/(2\lambda)} \rmd t \eqsp .
\end{equation}
Using \eqref{eq:vol_ball}, for all $k \geq 0$, $\int_{\rset_+}
t^{k}\rme^{t^2/(2\lambda) } \rmd t = (2\lambda)^{(k+1)/2}
\Gammabf((k+1)/2)$ and for all $a >1$, $\Gammabf(a+1/2) \leq a^{1/2}
\Gammabf(a)$ (by log-convexity of the Gamma function), we have
\begin{equation}\label{eq:B-bound-2}
C \leq \CUnfK^{-1} \sum_{i=0}^{d-1} \frac{\VolInt_i(\convSet)}{\vol(\convSet)} \parenthese{2\uppi\lambda}^{(d-i)/2}
\defEns{\rext + \left[\lambda(d-i+2)\right]^{1/2} } \eqsp .
\end{equation}
Regarding $D$ defined in \eqref{eq:def-A-B}, by \Cref{assumption:K}, \Cref{assumption:fK}-\ref{item:assum-fK-1}, \eqref{eq:indi_conv_proof1} and \eqref{eq:Wills}, we get:
\begin{equation}\label{eq:A-bound-1}
D \leq \rext \CUnfK^{-1} \DconstConvInt(\convSet,\lambdaMY) \eqsp,
\end{equation}
where $\DconstConvInt(\convSet,\lambdaMY)$ is defined in \eqref{eq:def-DKlambda}.
Combining \eqref{eq:B-bound-2} and \eqref{eq:A-bound-1} in \eqref{eq:vt-wasserstein}  concludes the proof.

\item 
Using \eqref{eq:InequalityVolInt} and \eqref{eq:MonoQuerm} in \eqref{eq:B-bound-2} gives for all $\lambda\in\ooint{0, (2\uppi)^{-1}\rin^2 d^{-2}}$
\begin{align*}\label{eq:B-bound-3}
C &\leq \CUnfK^{-1} \sum_{i=0}^{d-1} \parenthese{\frac{d}{\rin}\sqrt{\frac{\uppi\lambda}{2}}}^{d-i} \defEns{\rext + \left[\lambda(d-i+2)\right]^{1/2} } \\
& \leq \CUnfK^{-1} (2\uppi\lambda)^{1/2} d \rin^{-1} \parenthese{\rext+\rin\parenthese{\frac{3}{2d\uppi}}^{1/2}} \eqsp .
\end{align*}
Finally this bound,  \eqref{eq:A-bound-1}, \eqref{eq:boundDconst} and \eqref{eq:vt-wasserstein} conclude the proof.


\item
The proof still relies on the decomposition \eqref{eq:vt-wasserstein}, where $C$ and $D$ are defined in \eqref{eq:def-A-B}. Eq. \eqref{eq:WT-1} gives
\begin{equation*}
C \leq \int_{0}^{\plusinfty} (\rext+t) t\lambda^{-1} \rme^{-t^2/(2\lambda)} \parenthese{\frac{\int_{\convSet + \ball{0}{t}} \rme^{-f(x)} \rmd x}{\int_{\convSet} \rme^{-f(x)} \rmd x} -1} \rmd t \eqsp.
\end{equation*}
Under \Cref{assumption:fK}-\ref{item:assum-fK-2}, 
following the steps of \Cref{sec:proof-pi-pilambda-tv}-\ref{item:proof-TV-3} to upper bound the term $\int_{\convSet + \ball{0}{t}} \rme^{-f(x)} \rmd x / \int_{\convSet} \rme^{-f(x)} \rmd x$, we have
\begin{align*}
C &\leq \int_{0}^{\plusinfty} (\rext+t) t\lambda^{-1} \rme^{-t^2/(2\lambda)} \parenthese{(1+t/\rin)^d \rme^{(t\CDeuxfK)/\rin} -1} \rmd t \\
&= C_1 + C_2 \eqsp,
\end{align*}
where
\begin{align*}
C_1 &= \rext \int_{0}^{\plusinfty} t\lambda^{-1} \rme^{-t^2/(2\lambda)} \parenthese{(1+t/\rin)^d \rme^{(t\CDeuxfK)/\rin} -1} \rmd t \eqsp, \\
C_2 &= \int_{0}^{\plusinfty} t^2 \lambda^{-1} \rme^{-t^2/(2\lambda)} \parenthese{(1+t/\rin)^d \rme^{(t\CDeuxfK)/\rin} -1} \rmd t \eqsp.
\end{align*}
$C_1$ is upper bounded in the same way as $B$ in \Cref{sec:proof-pi-pilambda-tv}-\ref{item:proof-TV-3}. Regarding $C_2$, since for all $t \geq 0$, $(\CDeuxfK t)/\rin - t^2/(2\lambda) \leq -t^2/(4\lambda) + 4\lambda(\CDeuxfK/\rin)^2$, developping $(1+t/\rin)^d$ and using the change of variable $t\mapsto t^2/(4\lambda)$ we get
\begin{align*}
C_2 &\leq \rme^{4\lambda(\CDeuxfK/\rin)^2} \sum_{i=0}^d \binom{d}{i} \rin^{-i} \int_{\rset_+} t^{i+2} \lambda^{-1} \rme^{-t^2/(4\lambda)} \rmd t \\
&\leq 4\sqrt{\lambda} \rme^{4\lambda(\CDeuxfK/\rin)^2} \frac{\sqrt{\pi}}{2}\sum_{i=0}^d \binom{d}{i} \parenthese{\frac{2\sqrt{\lambda}}{\rin}}^i \Gammabf\parenthese{\frac{3}{2}+\frac{i}{2}} \eqsp.
\end{align*}
Using $\binom{d}{i}\Gammabf((3+i)/2) \leq (\sqrt{\pi}/2)d^i$ for $i\in\defEns{0,\ldots,d}$, we have for $\lambda\in\ocint{0,16^{-1}\rin^2d^{-2}}$,
\begin{align*}
C_2 &\leq 2\sqrt{\pi\lambda} \rme^{4\lambda(\CDeuxfK/\rin)^2} \sum_{i=0}^d \parenthese{\frac{2\sqrt{\lambda}d}{\rin}}^i \\
&\leq 4\sqrt{\pi\lambda} \rme^{4\lambda(\CDeuxfK/\rin)^2} \eqsp.
\end{align*}
$D$ defined in \eqref{eq:def-A-B} is upper bounded by $\rext B$ where $B$ is defined in \Cref{sec:proof-pi-pilambda-tv}-\ref{item:proof-TV-3}.
Combining the bounds on $C_1,C_2,D$ gives the result.
\end{enumerate}

\subsection{Proof of \Cref{prop:wasserstein-epsilon}}
\label{sec:proof-prop-wasserstein-epsilon}

Assume that $\gamma \in \ooint{0,(\m+\LUl)^{-1}}$. \cite[Theorem 5]{durmus:highdimULA2016} gives for all $n\in\nset^\star$:
\begin{equation*}
W_2^2(\delta_x \RKer_{\gaStep}^n, \pil) \leq 2 \left(1-(\kappa\gaStep)/2\right)^n \left\{ \norm[2]{x-\xstar}+d/\m\right\} + u(\gaStep) \eqsp,
\end{equation*}
where,
\begin{equation*}
u(\gaStep) = 2\kappa^{-1} \LUl^2 d \gaStep (\kappa^{-1}+\gaStep)\left( 2 + \frac{\LUl^2 \gaStep}{\m} + \frac{\LUl^2 \gaStep^2}{6} \right) \eqsp .
\end{equation*}
Noting that $\kappa \gamma \leq 1$ and $\LUl^2\gamma^2 \leq 1$, it is then sufficient for $\gamma,n$ to satisfy,
\begin{align*}
4\kappa^{-2} \LUl^2 d \gaStep \parenthese{ 2 + \frac{1}{6} + \frac{\LUl^2 \gaStep}{\m} } &\leq \varepsilon^2/2 \eqsp , \\
2 \left(1-(\kappa\gaStep)/2\right)^n \defEns{ \norm[2]{x-\xstar}+d/\m } &\leq \varepsilon^2/2 \eqsp ,
\end{align*}
which concludes the proof.

%
%
%


\section*{Acknowledgments} 

The authors wish to express their thanks to the anonymous referees for several helpful remarks, in particular concerning a simplified proof of \Cref{propo:finite-measure-MY}.

\def\sectionautorefname{Section}
\def\subsectionautorefname{Section}
\def\subsubsectionautorefname{Section}
\def\corollaryautorefname{Corollary}

 \printbibliography

\appendix

\section{Details of the orders of magnitude for \Cref{table:convergence-tvnorm} and \Cref{table:convergence-tvnorm-cunkf-cdeuxfk} }
\label{appendix:details-table-tv}

\begin{table}[!h]
\begin{center}
\begin{tabular}{|c|c|c|c|c|c|c|}
\hline 
  & $d \rightarrow \plusinfty $ & $\varepsilon \rightarrow 0$ & $\rext \rightarrow \plusinfty $ & $\rin \rightarrow 0$ & $\CUnfK \rightarrow 0$ & $\CDeuxfK \rightarrow \plusinfty$ \\ 
\hline 
$L, \lambda^{-1}$ & $d^2$ & $\varepsilon^{-2}$ & $1$ & $\rin^{-2}$ & $\CUnfK^{-2}$ & $\CDeuxfK^{2}$ \\ 
\hline 
$A_1(x)$ & $d^4$ & $\varepsilon^{-4}$ & $\rext^2$ & $\rin^{-4}$ & $\CUnfK^{-4}$ & $\CDeuxfK^{4}$\\
\hline
$-\log(\kappa)$ & $1$ & $1$ & $\rext^{-2}$ &  $1$ & $1$ & $1$ \\
\hline
$A_2(x)$ & $1$ & $\varepsilon^{-1}$ & $\rext$ &  $\rin^{-1}$ & $\CUnfK^{-1}$ & $\CDeuxfK$ \\
\hline
$T$ & $1$ & $\log(\varepsilon^{-1})$ & $\rext^2$ &  $\log(\rin^{-1})$ & $\log(\CUnfK^{-1})$ & $\log(\CDeuxfK)$ \\
\hline
$\gamma$ & $d^{-5}$ & $\varepsilon^{6}$ & $\rext^{-2}$ &  $\rin^{-4}$ & $\CUnfK^{4}$ & $\CDeuxfK^{-4}$ \\
\hline
\end{tabular} 
\end{center}
\caption{dependency of $L, A_1(x), -\log(\kappa), A_2(x), T, \gamma$ on $d$, $\varepsilon$, $\rext$, $\rin$, $\CUnfK$ and $\CDeuxfK$.}
\label{table:convergence-tvnorm-details}
\end{table}

\end{document}